\documentclass[a4paper,11pt]{article}
\pdfoutput=1

\usepackage{jcappub}

\usepackage[T1]{fontenc} 
\usepackage{framed}
\usepackage{multirow}
\usepackage{subcaption}
\newcommand{\ignore}[1]{}

\title{\boldmath $Sp(4,\mathbb{Z})$ modular inflation}

\author[a,1]{Si-Yi Jiang,\note{Corresponding author}}
\author[b,c]{Wenbin Zhao\,,}
\author[a,d]{Gui-Jun Ding\,}

\affiliation[a]{Department of Modern Physics,  and Anhui Center for fundamental sciences in theoretical physics, University of Science and Technology of China, Hefei, Anhui 230026, China}
\affiliation[b]{International Centre for Theoretical Physics Asia-Pacific, University of Chinese Academy of Sciences,
100190 Beĳing, China}
\affiliation[c]{Taiji Laboratory for Gravitational Wave Universe (Beijing/Hangzhou), University of Chinese Academy of Sciences (UCAS), Beijing, China.}
\affiliation[d]{College of Physics, Guizhou University, Guiyang 550025, China}

\emailAdd{siichiang@mail.ustc.edu.cn}
\emailAdd{zhaowenbin@ucas.ac.cn}
\emailAdd{dinggj@ustc.edu.cn}

\abstract{
We investigate inflation models governed by the Siegel modular group $Sp(4,\mathbb{Z})$.
The $Sp(4,\mathbb{Z})$ group extends the $SL(2,\mathbb{Z})$ framework from one modulus to three moduli while preserving the hyperbolic geometry of the K\"ahler potential, allowing for the construction of cosmological $\alpha$-attractor models.
In this context, we use genus $g=2$ absolute invariants to construct inflationary potentials within specific subspaces of the Siegel moduli space. These models are driven by the imaginary components of the moduli $\tau$ and naturally yield plateau-like potentials consistent with Planck 2018 observations in large field limit. We employ two-dimensional complex subspaces to realize E-model and T-model like two-field inflation scenarios. We explore the subspace of complex dimension one to construct a modified polynomial $\alpha$-attractor model, which can accommodate the larger spectral index $n_s$ favored by recent ACT and SPT data, particularly in the larger $N$ regime.
}

\begin{document}
\maketitle
\flushbottom

\section{Introduction\label{sec:intro}}
Inflation provides an elegant mechanism to resolve the flatness, horizon, and monopole problems in the early evolution of the universe~\cite{Starobinsky:1980te,Guth:1980zm,Linde:1981mu,Albrecht:1982wi}. Moreover, precise measurements of the cosmic microwave background (CMB) strongly support the predictions of inflationary models~\cite{Planck:2018jri,BICEP2:2018kqh,ACT:2025fju,ACT:2025tim,SPT-3G:2025bzu}. Recently, it is remarkable that the
modulus field $\tau$ could be identified as inflaton to realize both large and small field inflation within a $SL(2,\mathbb{Z})$ modular invariant effective potential, while modular invariance imposes strong constraints on the scalar potential of the moduli fields~\cite{Schimmrigk:2014ica,Schimmrigk:2015qju,Schimmrigk:2016bde}. In terms of large field inflation, Starobinsky-like inflationary models can naturally emerge within the framework of modular symmetry~\cite{Casas:2024jbw}. Linde and his collaborators observed that $\alpha$-attractor inflation models could be naturally realized from modular invariance, specifically the T-model and E-model, owing to the asymptotic properties of the modular functions $j$ and $\eta$~\cite{Kallosh:2024ymt,Kallosh:2024pat,Kallosh:2024kgt}.
On the other hand, based on the properties of the potential at its fixed points $\tau=i\,,\omega$, small field inflation can be naturally realized if the modulus slowly rolls down along the unit arc boundary of the $SL(2,\mathbb{Z})$ fundamental domain~\cite{Ding:2024neh,King:2024ssx}.
Modular symmetry provides a unifying framework to simultaneously address the lepton flavor puzzle, inflation, and the reheating of the Universe~\cite{Ding:2024euc}. There are also other scenarios of the modular inflation proposed~\cite{Abe:2023ylh,Kobayashi:2016mzg}, which employed a stabilizer field coupled to a modular form. The scalar potential is flattened to realize inflation via corrections to the K\"ahler potential or scalar potential. The Higgs-modular inflation model extends modular inflation by coupling the Higgs boson to enhance the reheating temperature and yield predictions compatible with ACT observations~\cite{Aoki:2025wld}.

Inflation is often assumed to be driven by a single scalar field under the slow-roll approximation. These models predict a nearly scale-invariant and Gaussian-distributed spectrum of primordial density perturbations. However, supersymmetric field theories and string theory generally give rise to multiple scalar fields, which may actively participate in the dynamics of inflation~\cite{Adams:1997de,Wands:2007bd,Gong:2016qmq}. The multi-field inflation allows for the exploration of richer phenomenology, such as the evolution of curvature and isocurvature perturbations, as well as the generation of non-Gaussian fluctuations~\cite{Wands:2007bd,Seery:2005gb}.

The Siegel modular symmetry $Sp(2g,\mathbb{Z})$ is a natural generalization of the
well-known $SL(2,\mathbb{Z})$ modular symmetry, and the corresponding moduli $\tau$ is represented by a complex $g\times g$ matrix with positive definite imaginary part, where $g$ is called genus. For $g=1$, it coincide the $SL(2,\mathbb{Z})$ modular group exactly with $Sp(2,\mathbb{Z})\cong SL(2,\mathbb{Z})$. The Siegel modular symmetry
naturally arise from string theory compactifications of extra  dimensions~\cite{alma9938900631702711,Bailin:1998yt,LopesCardoso:1996nc}. Siegel modular invariance has also been extensively applied in the construction of modular flavor models, where the symmetry severely constrains the structure of Yukawa couplings through modular forms, thereby providing a natural framework to address the flavor puzzle~\cite{Ding:2020zxw,Ding:2024xhz,Baur:2020yjl,Nilles:2021glx}. The stabilization of multiple moduli within the framework of $Sp(4,\mathbb{Z})$ symmetry has also been actively studied~\cite{Funakoshi:2024yxg}. In light that the $\alpha$-attractor models involve single inflation field can be realized from $SL(2,\mathbb{Z})$ symmetry, we expect the $\alpha$-attractor models with multi fields can be obtained from Siegel modular group, where the multiple moduli play role of inflatons.

The $\alpha$-attractor models were originally proposed to unify a broad class of inflationary scenarios whose predictions converge to the universal attractor regime with~\cite{Kallosh:2013yoa,Kallosh:2013hoa,Kallosh:2013daa,Ferrara:2013rsa,Kallosh:2013maa,Kallosh:2013tua}:
\begin{equation}\label{eq:alpnsr}
n_s = 1 - \frac{2}{N}\,, \quad\quad  r = \frac{12\alpha}{N^2}\,,
\end{equation}
where the parameter $\alpha$ controls the curvature of the inflaton's K\"ahler manifold and determines the tensor-to-scalar ratio $r$. For $g=1$, under the $SL(2,\mathbb{Z})$ symmetry, the geometric origin of $\alpha$-attractor lies in their K\"ahler potential~\cite{Kallosh:2013yoa,Kallosh:2015zsa,Achucarro:2017ing}:
\begin{equation}
\label{eq:Kahler-potential}{\cal K} = -3\alpha \log\left[-i(\tau-\bar{\tau})\right]\,,
\end{equation}
where $\tau$ is inflaton superfield, and the curvature of the K\"ahler manifold is given by
\begin{equation}
{\cal R}_{\text{K\"ahler}}= -{\cal K}_{\tau\bar{\tau}}^{-1}\partial_\tau \partial_{\bar{\tau}}\log ({\cal K}_{\tau\bar{\tau}})= -\frac{2}{3\alpha}\,,
\end{equation}
which is a hyperbolic geometry and it is helpful to construct $\alpha$-attractor models. After canonically normalizing the kinetic term for the imaginary part of $\tau$, the modular invariant scalar potential could develop an exponentially flat plateau so that slow-roll inflation could be naturally realized.

In the present work, we consider the case of $g=2$ and the Siegel group $Sp(4,\mathbb{Z})$. The K\"ahler potential takes the following form~\cite{Ding:2020zxw}:
\begin{equation}
\label{eq:mtauKp}
{\cal K} = - 3\alpha\log\det[-i(\tau-\bar{\tau})] = - 3\alpha \log\left[(\tau_3-\bar{\tau}_3)^2 - (\tau_1-\bar{\tau}_1)(\tau_2-\bar{\tau}_2)\right]\,,
\end{equation}
where $\tau=
\begin{pmatrix}
\tau_1~&~ \tau_3\\
\tau_3~&~ \tau_2
\end{pmatrix}
$ is a two-by-two complex symmetric matrix and $\alpha$ is a positive constant. One sees that there are three complex moduli $\tau_1$, $\tau_2$ and $\tau_3$.
The Ricci scalar of the corresponding K\"ahler geometry is given by~\cite{Freedman:2012zz}:
\begin{equation}
{\cal R}_{\text{K\"ahler}} =-{\cal K}^{\alpha\bar{\beta}}\partial_{\alpha}\partial_{\bar{\beta}}\log(\det {\cal K}_{\rho\bar{\gamma}})\,,
\end{equation}
where ${\cal K}^{\alpha\bar{\beta}}$ is inverse of the K\"ahler metric ${\cal K}_{\alpha\bar{\beta}}=\partial_\alpha\partial_{\bar{\beta}} {\cal K}$. Then we can determine the Ricci scalar to be
\begin{equation}
{\cal R}_{\text{K\"ahler}} = -\frac{3}{\alpha}\,,
\end{equation}
which confirms the hyperbolic nature of the field space. This geometric similarity suggests that models could naturally  generalize the $\alpha$-attractor inflation from single modulus to multiple moduli.

The kinetic terms of moduli are determined by the K\"ahler potential.
To study the dynamics of  inflation driven by the moduli fields, we also need to specify the scalar potentials which are invariant under the Siegel modular transformation.
In analogy with modular inflation models based on $SL(2,\mathbb{Z})$, where the potential is built from the Klein $j$-invariant~\cite{Kallosh:2024ymt,Kallosh:2024pat,Kallosh:2024kgt}.
In the context of Siegel modular symmetry for $g=2$, there are three absolute invariants $y_1$, $y_2$, and $y_3$ which are expressed by the Siegel modular forms $\psi_4$, $\psi_6$, together with the cusp forms $\chi_{10}$ and $\chi_{12}$~\cite{Igusa1967ModularFA,6c36a4a1-df00-35e4-8934-01b06a1026c9}.
These invariants serve as the building blocks for constructing scalar potentials that realize multi-field $\alpha$-attractor models.

This paper is organized as follows. We begin in section~\ref{sec:Muti-frame} by establishing the general framework of multi-field inflation, reviewing the relevant slow-roll parameters and observational predictions. Section~\ref{sec:SMS} lays the mathematical foundation, introducing the key concepts of Siegel modular invariance with a focus on Siegel modular forms and absolute invariants. The core of our analysis is presented in section~\ref{sec:Model}, where we construct specific inflationary models across three distinct subspaces. First, subsection~\ref{sec:tau12} explores the factorized subspace $\tau=\text{diag}(\tau_1, \tau_2)$, and the absolute invariant $y_2(\tau)$ is used to construct modular invariant scalar potential. Next, the modular subspace
$\tau=\begin{pmatrix}
\tau_1~&~\tau_3\\
\tau_3~&~\tau_1
\end{pmatrix}$ is studied in subsection~\ref{sec:tau13}, the modulus field $\tau_1$ and $\tau_3$ couple with each other in both K\"ahler potential and absolute invariants. We implement field redefinition by rotating the complex variables, thereby realizing the inflation by using the absolute invariant $y_3(\tau)$. Finally, subsection~\ref{sec:tau11} investigates the single modulus subspace
$\tau=\begin{pmatrix}
\tau_1~&~\tau_1/2\\
\tau_1/2~&~\tau_1
\end{pmatrix}$, where we construct a polynomial $\alpha$-attractor model based on $y_1(\tau)$ that reproduces the spectral index favored by recent ACT~\cite{ACT:2025fju,ACT:2025tim} and SPT~\cite{SPT-3G:2025bzu} observations. We conclude with a discussion of our results in section~\ref{sec:dis}.

\section{Framework of multi-field inflation \label{sec:Muti-frame}}

The general action for $n$ real scalar inflaton fields $\phi^a(a=1,\ldots,n)$ coupled with gravity takes the following form
\begin{equation}\label{eq:SactL}
S=\int \text{d}^4 x \sqrt{-g}\left[\frac{M_{\text{Pl}}^2}{2}R + \frac{1}{2}g^{\mu\nu}{\cal K}_{ab}\partial_\mu\phi^a\partial_\nu\phi^b-V(\phi^a)\right]\,,
\end{equation}
where $M_{\text{Pl}}=\frac{1}{\sqrt{8\pi G}}$ is the reduced Planck constant and is denoted $1$ for convention, $R$ is the Ricci scalar, ${\cal K}_{ab}\equiv {\cal K}_{ab}(\phi^c)$ denotes the K\"ahler metric for the real scalar fields, and $V(\phi^a)$ is the scalar potential. In a spatially flat, homogeneous, and isotropic FLRW universe, the metric and the inflaton field take the form:
\begin{equation}
ds^2 = dt^2 - a^2(t)\, d\vec{x}^2\,, \qquad
\phi^a(t, \vec{x}) = \phi^a_0(t)\,.
\end{equation}
The Friedmann equation is given by
\begin{equation}\label{eq:FeqKG}
H^2 = \frac{1}{3 }\left( \frac{1}{2}{\cal K}_{ab}\dot{\phi}_0^{a}\dot{\phi}_0^{b} + V\right)=\frac{1}{3 }\left(\frac{1}{2}\dot{\phi}^2_0 + V\right) \,,
\end{equation}
while the equation of motion reads
\begin{equation}
\label{eq:FeqKGf}
D_t\dot{\phi}_0^a + 3H \dot{\phi}_0^a + {\cal K}^{ab} \frac{\partial V}{\partial \phi^b_0} =0\,,
\end{equation}
where ${\cal K}^{ab}$ is the inverse of the K\"ahler metric ${\cal K}_{ab}$, $\dot{\phi}_0 \equiv \sqrt{{\cal K}_{ab}\dot{\phi}_0^{a}\dot{\phi}_0^{b}}$ and $D_t$ acting on a generic vector $v^a$ in field space in the following way:
\begin{equation}\label{eq:Dnabla}
D_t v^a\equiv \dot{\phi}_0^b \nabla_b v^a  =  \dot{v}^{a} + \gamma^a_{bc}v^c \dot{\phi}_0^b\,,
\end{equation}
with
\begin{equation}
\nabla_b v^a = \frac{d v^a}{d \phi_0^b} +  \gamma^a_{bc}v^c\,,\quad
\nabla_b v_a = \frac{d v_a}{d \phi_0^b} -  \gamma^c_{ab}v_c\,,
\end{equation}
here $\gamma^a_{bc}$ and $\nabla_a$ are the Levi-Civita connection and
the covariant derivative respectively. The connection $\gamma^a_{bc}$ is defined as~\cite{Sasaki:1995aw}
\begin{equation}\label{eq:gammadef}
\gamma^a_{bc}\equiv \frac{1}{2}{\cal K}^{ad} \left({\cal K}_{db,c} + {\cal K}_{dc,b} - {\cal K}_{bc,d}\right)\,,
\end{equation}
where $,a$ denotes the derivative with respect to $\phi_0^a$.
Moreover, the derivative of kinetic term  with respect to $t$ satisfy the following relation
\begin{equation}\label{eq:kin}
\frac{1}{2}\frac{d}{dt}\dot{\phi}_0^2=\frac{d}{dt}\left(\frac{1}{2}{\cal K}_{ab}\dot{\phi}_0^{a}\dot{\phi}_0^{b}\right)={\cal K}_{ab}(D_t\dot{\phi}_0^{a})\dot{\phi}_0^{b}\,,
\end{equation}
which is useful for calculation in multi-field perturbation.

The slow-roll conditions in multi-field inflation can be understood neatly using a kinematic basis to decompose the inflationary trajectory into tangent and normal directions (i.e. adiabatic and entropic). Focusing on
the two field case for concreteness\footnote{The general formula of multiple field inflation are introduced in section~3.1 of \cite{GrootNibbelink:2000vx}.}, we introduce unit tangent (adiabatic) and normal (entropic) vectors $\sigma^a$ and $s^a$ as follow~\cite{Cicoli:2023opf}
\begin{equation}\label{eq:TNdir}
\sigma^a \equiv \frac{\dot{\phi}^a_0}{\dot{\phi}_0}\,,\quad \sigma^a \sigma_a =1\,,\quad s^a \sigma_a =0\,,\quad s^a s_a =1\,.
\end{equation}
In field space, $\sigma^a$ is parallel to the inflationary trajectory in field space, while $s^a$ is the orthogonal to the trajectory. Since $s^a$ is naturally proportional to the derivative of $\sigma^a$, i.e. $D_t \sigma^a \propto s^a$, we define $s^a$ as~\cite{Achucarro:2010da}
\begin{equation}
s^a\equiv \left({\cal K}_{bc}D_t\sigma^b D_t\sigma^c\right)^{-1/2}D_t\sigma^a \,,
\end{equation}
where $D_t\sigma^a$ can be written as by using Eq.~\eqref{eq:FeqKGf}
\begin{equation}
D_t \sigma^a = -\frac{\ddot{\phi}_0}{\dot{\phi}_0}\sigma^a - \frac{1}{\dot{\phi}_0}\left(3H\dot{\phi}_0^a + V^{,a}\right)\,.
\end{equation}
Decomposing this equation along the orthogonal basis vectors $\sigma^a$ and $s^a$, we find
\begin{eqnarray}
\label{eq:phi0KG} \ddot{\phi}_0 + 3 H\dot{\phi}_0 &+& V_\sigma  = 0\,,\\
\label{eq:DtTafinal}
D_t \sigma^a
&=& - \frac{V_s}{\dot{\phi}_0}s^a\,,
\end{eqnarray}
where we denote $V_\sigma\equiv \sigma^a V_{,a}$ and $V_s\equiv s^a V_{,a}$, corresponding to the components of $V_{,a}\equiv \frac{\partial V}{\partial \phi_0^a}$ along the tangent and normal directions, respectively.

Differentiating Eq.~\eqref{eq:FeqKG} with respect to time $t$ and using Eq.~\eqref{eq:phi0KG}, we can obtain
\begin{equation}\label{eq:Hdot-eq}
\dot{H} = - \frac{\dot{\phi}_0^2}{2 }\,,
\end{equation}
which is useful for calculating Eq.~\eqref{eq:phiadotapprox} and Eq.~\eqref{eq:Na}.

\subsection{The slow-roll approximation}
To study slow-roll inflation, it is useful to define the following
dimensionless quantities~\cite{Achucarro:2010da}:
\begin{eqnarray}
\label{eq:ep1}\epsilon_H &\equiv& -\frac{\dot{H}}{H^2} = \frac{\dot{\phi}_0^ 2}{2 H^2}\,,\\
\label{eq:eta1}\eta^a &\equiv& -\frac{1}{H\dot{\phi}_0} D_t\dot{\phi}^a_0\,.
\end{eqnarray}
The accelerating expansion of Universe requires $\epsilon_H<1$.
The parameter $\eta^a$ is introduced to describe the ratio between the acceleration of scalar field and friction term. The vector $\eta^a$ can be decomposed of $\sigma^a$ and $s^a$ as
\begin{eqnarray}
\eta^a = \eta_{\parallel}\,\sigma^a + \eta_{\perp}\, s^a\,,\quad \quad\quad\eta_{\parallel} = -\frac{\ddot{\phi}_0}{H\dot{\phi}_0}\,,\quad \eta_{\perp} &=&\frac{V_s}{\dot{\phi}_0 H }\,,
\end{eqnarray}
where $\eta_{\parallel}$ depends only on the direction of the inflaton field.
To achieve a sufficient number of $e$-folds of inflation, the parameter $\epsilon_H$ must remain much smaller than $1$ in inflation period. Consequently, the parameter quantifying the change rate of $\epsilon_H$ is defined as
\begin{equation}
\label{eq:slowroll2single}
\eta_H \equiv H^{-1}\frac{ \dot{\epsilon}_H}{\epsilon_H} \,.
\end{equation}
It is straightforward to find that $\eta_H$ can be expressed in terms of $\epsilon_H$ and $\eta_{\parallel}$ as follow
\begin{equation}
\eta_H = 2\left( \epsilon_H - \eta_{\parallel}\right)\,.
\end{equation}
Under slow-roll approximation,
$\epsilon_H\ll1$ and $\eta_H\ll 1$ are satisfied, we find
\begin{equation}
\eta_{\parallel} \ll 1\,,
\end{equation}
which implies that   Eq.~\eqref{eq:FeqKG} and Eq.~\eqref{eq:phi0KG} can be approximately written as
\begin{equation}\label{eq:approx}
3 H\dot{\phi}_0 \approx - V_\sigma \,,\quad H^2 \approx \frac{V}{3}\,.
\end{equation}
Moreover, by using Eq.~\eqref{eq:Hdot-eq} and Eq.~\eqref{eq:approx}, the $\dot{\phi}_0^a$ can be expressed in terms of the scalar potential
\begin{equation}
\label{eq:phiadotapprox}  \dot{\phi}_{0}^a \approx - \frac{  V^{,a}}{\sqrt{3 V}}\,.
\end{equation}
Based on Eq.~\eqref{eq:phiadotapprox}, the covariant time derivative can also be expressed in terms of the covariant derivative with respect to the fields and derivatives of the scalar potential.
\begin{equation}\label{eq:DtV}
D_t =
\dot{\phi}_0^a \nabla_a  \approx -\frac{V^{,a}}{ \sqrt{3 V}} \nabla_a\,.
\end{equation}
Equipped with these relations under slow-roll approximation,
one can express the the slow-roll conditions as constraints on the scalar potential $V$, and we define
\begin{equation}\label{eq:epVoprep}
\epsilon_V\equiv\frac{1}{2}\frac{ V^{,a}V_{,a}}{ V^2 } \approx \epsilon_H \,.
\end{equation}
Similarly,  $\eta_{\parallel}$ can be written as
\begin{equation}\label{eq:evetaV}
\eta_{\parallel} = \eta_V{}^{a}_{b}\frac{V_{,a}V^{,b}}{V^{,c}V_{,c}} - \epsilon_V\,,\quad \eta_V{}^{a}_{b}\equiv \frac{V^{;a}_{;b}}{V}\,,
\end{equation}
where $;a$ denotes covariant derivative with respect to $\phi_0^a$.
The second slow-roll parameter $\eta_H$ in Eq.~\eqref{eq:slowroll2single} can also be approximated in terms of the potential slow-roll parameters
\begin{eqnarray}
\eta_H= 4\epsilon_V-2\eta_V{}^{a}_{b}\frac{ V^{,b}V_{,a}}{ V^c V_c } \ll 1\,.
\end{eqnarray}
It is instructive to present an explicit expression for the number of e-foldings $N$. In multi-field inflation, the inflaton trajectory uniquely determines $N$, and its derivative with respect to the field coordinates is given by
\begin{equation}\label{eq:Na}
N_{,a} =  -\frac{V V_{,a}}{V^{,b}V_{,b}}\,.
\end{equation}
The inflationary observables should be calculated at the point of  horizon exit, which is defined by the pivot scale $k_* = a H $ that represents the scales measured by the CMB. In two-field inflation scenario, the field values at this moment are denoted as $(\varphi_{1*},\varphi_{2*})$. The number of e-folds of inflation, $N$, is given by
\begin{equation}
N(\varphi_{1*},\varphi_{2*}) = \int_{(\varphi_{1*},\varphi_{2*})}^{(\varphi_{1\text{end}},\varphi_{2\text{end}})}N_{,a}d\varphi^a\,,
\end{equation}
where $\varphi_{1\text{end}}$ and $\varphi_{2\text{end}}$ denote the field values in the end of inflation.

\subsection{Predictions of slow-roll multi-field inflation}
\label{subsec:sr-multifield}

In slow-roll multi-field inflation, the observed scalar and tensor spectra and their mild scale dependence are determined by the dynamics of multiple interacting fields. The key observables include the spectral index $n_s$, the tensor-to-scalar ratio $r$, and the running of the spectral index $\alpha_s$. These quantities can be written in terms of the slow-roll parameters determined by the potential $V(\phi^a)$ and its covariant derivatives on field space, as summarized below. To connect with observations, predictions for non-Gaussianity from the three-point function can be expressed in terms of derivatives of $N$.

\subsubsection*{Curvature perturbations and the power spectrum}
\label{subsubsec:scalar-ps}

In the baseline $\Lambda$CDM framework, the initial state of perturbations is taken to be purely adiabatic and (to leading order) Gaussian. The spectrum of curvature perturbations produced during inflation is of the following form~\cite{Sasaki:1995aw}
\begin{equation}
\label{eq:PR}
\mathcal{P}_{\mathcal{R}} = A_s \left(\frac{k}{k_{*}}\right)^{n_s-1}\,,\quad  A_s=\left(\frac{H}{2\pi}\right)^{\!2} N^{,a} N_{,a}\,,
\end{equation}
where $k_{*}$ is pivot scale and $A_s$ is amplitude of scalar spectrum.
The spectral index of the curvature perturbations is~\cite{Sasaki:1995aw}
\begin{eqnarray}
\label{eq:ns-deltaN}
\nonumber n_s - 1\equiv \frac{\text{d}\ln \mathcal{P}_{\mathcal{R}}}{\text{d}\ln k} \;&=&\; -2\epsilon_V \;-\; \frac{1}{N^{,a}N_{,a}}
\;+\; 2\,\left(\eta_V\right)^a_{\;b}\,\frac{N_{,a}N^{,b}}{N^{,c}N_{,c}}\\
\;&=&\; -3  \,\frac{V^{,a}V_{,a}}{V^2}
\;+\; 2\,\left(\eta_V\right)^a_{\;b}\,\frac{V_{,a}V^{,b}}{V^{,c}V_{,c}}\,.
\end{eqnarray}

\subsubsection*{Running of the spectral index}
\label{subsubsec:running}

The running of the spectral index $\alpha_s \equiv \mathrm{d}n_s/\mathrm{d}\ln k$ quantifies mild scale dependence beyond a pure power law. In slow-roll multi-field inflation, it can be written in terms of the potential and the covariant derivatives as follow,
\begin{eqnarray}
\label{eq:alphas-def}
\nonumber \alpha_s \;&=&\; 8 \,\left(\eta_V\right)^{a}_{\;b}\,\frac{V^{,b}V_{,a}}{V^2}
\;-\; 24\,\epsilon_V^2
\;-\; 2\,\left(\xi_V^2\right)^{a}_{\;b}\,\frac{V_{,a}V^{,b}}{V^{,c}V_{,c}}\\
\;&&-\; 4\,\eta_{V\,ab}\!\left[
\left(\eta_V\right)^{a}_{\;d}\,\frac{V^{,d}V^{,b}}{V^{,c}V_{,c}}
\;-\;
\left(\eta_V\right)^{c}_{\;d}\,\frac{V^{,a}V^{,b}V^{,d}V_{,c}}{(V^{,e}V_{,e})^2}
\right]\!,
\end{eqnarray}
where
\begin{equation}
\label{eq:xidef}
\left(\xi_V^2\right)^{a}_{\;b}\;\equiv\; \,\frac{V^{,d}\nabla_d \nabla^a \nabla_b V}{V^2}\,,
\qquad
\left(\eta_{V}\right)_{ab} \;\equiv\; \mathcal{K}_{ac}\,\left(\eta_V\right)^{c}_{\;b}\,.
\end{equation}
In slow-roll scenarios $\alpha_s$ is second order in slow-roll parameters and hence expected to be small. In the limit of single field inflation, both $\alpha_s$ and $n_s$ are reduced to well-known formula~\cite{Lyth:2009imm}.

Planck 2018, combined with baryon acoustic oscillation (BAO) measurements at the pivot scale $k_{*}=0.05 \text{Mpc}^{-1}$, give~\cite{Planck:2018vyg}
\begin{equation}
n_s = 0.9659 \pm 0.0040\,,\quad \alpha_s = -0.0041 \pm 0.0067\,.
\end{equation}

\subsubsection*{Tensor-to-scalar ratio }
\label{subsubsec:tensor-ps}

The rapid accelerated expansion during inflation generates nearly scale-invariant primordial tensor fluctuations on super-Hubble scales. These fluctuations give rise to a stochastic background of primordial gravitational waves in the post-inflationary epochs. The amplitude of this signal is usually quantified by the tensor-to-scalar ratio $r$, defined as the ratio of the tensor to scalar power spectrum amplitudes, and in slow-roll inflation it is characterized by the parameter $\epsilon_V$:
\begin{equation}
\label{eq:r-deltaN}
r \;=\; 16\,\epsilon_V\,.
\end{equation}
The most stringent bound on tensor-to-scalar ratio is given by BICEP/Keck 2018~\cite{BICEP:2021xfz}:
\begin{equation}\label{BK2018}
r < 0.036\,\quad \text{at 95\% \text{C.L.}}~~\,.
\end{equation}

\subsubsection*{Non-Gaussianity}

To analyze higher-order statistics it is convenient to adopt the $\delta N$ formalism, which relates the curvature perturbation $\zeta$ on uniform-density hypersurfaces to the local e-fold number $N$~\cite{Seery:2005gb}:
\begin{equation}
\label{eq:zeta-dN}
\zeta\;\equiv\; \delta N \;=\; N_{,a}\,\delta\phi^a \;+\; \frac{1}{2} N_{,b;a}\,\delta\phi^a\delta\phi^b \;+\; \cdots\,,
\end{equation}
where $\delta N$ is the local e-folding number between an initial flat hypersurface and a final uniform-density hypersurface, and $\delta\phi^a$ are the field fluctuations at horizon exit.
The bispectrum is defined via the three-point correlation function with
\begin{equation}
\langle \zeta_{\mathbf{k}} \zeta_{\mathbf{k}'} \zeta_{\mathbf{k}''} \rangle
\;\equiv\; (2\pi)^3 \delta^{(3)}(\mathbf{k}+\mathbf{k}'+\mathbf{k}'')\,B_\zeta(k,k',k'')\,,
\end{equation}
where $k$, $k'$ and $k''$ are magnitudes of the momenta, as dictated by the isotropy of the background, and the bispectrum in the local type is given by~\cite{Byrnes:2010ft}
\begin{equation}
B_\zeta(k,k',k'') \;=\;\frac{6}{5}\,f_{\rm NL}(k,k',k'') \,\big[ P_\zeta(k)P_\zeta(k') + P_\zeta(k')P_\zeta(k'') + P_\zeta(k)P_\zeta(k'') \big]\,,
\end{equation}
where $f_{\rm NL}$ is a non-linearity parameter and power spectrum $P_\zeta(k)$ is defined by the two-point correlation function $\langle \zeta_{\mathbf{k}} \zeta_{\mathbf{k}'} \rangle = (2\pi)^3 \delta^{(3)}(\mathbf{k}+\mathbf{k}') P_\zeta(k)$. The signal for this bispectrum is peaked in the squeezed limit $k\ll k'\approx k''$. Neglecting intrinsic non-Gaussianity of the horizon-exit field fluctuations $\delta\phi^a$, the non-linearity parameter can be written in a covariant form as
\begin{equation}
f_{\rm NL} \;\approx\; \frac{5}{6}\,\frac{N_{,b;a}\,N^{,a} N^{,b}}{\left(\mathcal{K}^{cd} N_{,c} N_{,d}\right)^2}\,.
\end{equation}
Current observations from Planck are consistent with $f_{\rm NL}$ close to zero~\cite{Planck:2019kim}
\begin{equation}
f_{\rm NL} \;=\; -0.9 \pm 5.1 \quad \text{at} \quad 68\%~\mathrm{CL}\,.
\end{equation}

\section{Siegel Modular symmetry\label{sec:SMS}}

The Siegel modular group $Sp(2g,\mathbb{Z})$ of genus $g$, is the group of $2g\times 2g$ integer matrices preserving a skew-symmetric bilinear form $J$, i.e.
\begin{equation}
Sp(2g,\mathbb{Z}) = \Big\{ \textsf{g} \in GL(2g,\mathbb{Z})~|~ \textsf{g}^t J \textsf{g} = J\Big\}\,,\quad \textsf{g}=
\left(
\begin{array}{cc}
A~&~B\\
C~&~D
\end{array}
\right)\,,\quad J=
\left(
\begin{array}{cc}
0~&~\mathbf{1}_g\\
-\mathbf{1}_g~&~0
\end{array}
\right)\,,
\end{equation}
where $\mathbf{1}_g$ represents the $g$-dimensional identity matrix. $A$, $B$, $C$ and $D$ are $g\times g$ integer matrices and they satisfy the following relations:
\begin{equation}
A^t C=C^t A\,,\qquad B^t D=D^t B\,,\qquad A^t D-C^t B=\mathbf{1}_g\,.
\label{p2}
\end{equation}
From this condition, one sees that $Sp(2g,\mathbb{Z})$ coincides with $SL(2,\mathbb{Z})$ when $g=1$. The Siegel modular group $Sp(2g,\mathbb{Z})$ acts on the moduli $\tau$ which is a symmetric complex $g\times g$ matrix with positive definite imaginary part, i.e.
\begin{equation}
\tau^t = \tau ,~~~~ \Im (\tau) > 0 \,.
\end{equation}
Similar to the case of $SL(2,\mathbb{Z})$, the Siegel modular transformation of $\tau$ is
\begin{equation}\label{eq:SiegelTrans}
\tau\to \textsf{g} \tau=(A \tau+B)(C\tau +D)^{-1}\,.
\end{equation}
The Siegel modular group $Sp(2g,\mathbb{Z})$ has infinite numbers of group elements, while it can be generated by the following set of matrices,
\begin{equation}
\label{genSieg}
{\cal S}=\left(
\begin{array}{cc}
0~&~\mathbf{1}_g\\
-\mathbf{1}_g~&~0
\end{array}
\right)~~~,~~~~{\cal T}_i=\left(
\begin{array}{cc}
\mathbf{1}_g~&~B_i\\
0~&~\mathbf{1}_g
\end{array}
\right)~~~,
\end{equation}
where $\{B_i\}$ with $i=1,2,\ldots,2g-1$ is a basis for the $g\times g$ integer symmetric matrices, and $\mathcal{S}$ coincides with the invariant symplectic form $J$ satisfying ${\cal S}^2=-\mathbf{1}_{2g}$.
Siegel modular forms $Y(\tau)$ of integer weight $k$ at genus $g$ are holomorphic functions that transform as
\begin{equation}
Y(\textsf{g}\tau) = \text{det}(C\tau+D)^{k} Y(\tau)\,,
\end{equation}
for all $\textsf{g}\in Sp(2g,\mathbb{Z})$. For genus $g=1$, this transformation law reduces to that of modular forms under $SL(2,\mathbb{Z})$. In this work, we focus on the case $g=2$, where $\tau$ is characterized by three moduli,
\begin{equation}\label{eq:totalspace}
\tau =
\begin{pmatrix}
\tau_1 ~&~ \tau_3 \\
\tau_3 ~&~ \tau_2
\end{pmatrix}\,,
\end{equation}
the condition of positive definite imaginary part of $\tau$ requires
\begin{equation}
\Im(\tau_1)\Im(\tau_2) - \Im(\tau_3)^2 > 0\,,\quad
\Im(\tau_1) + \Im(\tau_2)>0\,.
\end{equation}
In particular, for genus $g = 2$, the basis $\{B_i\}$ associated with the generators ${\cal T}_{1,2,3}$ is given by by~\cite{Ding:2020zxw}
\begin{equation}
B_1 =
\begin{pmatrix}
1 ~&~ 0\\
0 ~&~ 0
\end{pmatrix}\,,\quad
B_2 =
\begin{pmatrix}
0 ~&~ 0\\
0 ~&~ 1
\end{pmatrix}\,,\quad
B_3 =
\begin{pmatrix}
0 ~&~ 1\\
1 ~&~ 0
\end{pmatrix}\,.
\end{equation}

For simplicity, we shall study inflationary trajectories restricted to particular subspaces of the moduli space~\cite{Ding:2020zxw}
\begin{equation}\label{eq:subspace123}
\begin{pmatrix}
\tau_1 ~&~ \tau_1/2\\
\tau_1/2 ~&~ \tau_1
\end{pmatrix}\,,\quad
\begin{pmatrix}
\tau_1 ~&~ 0\\
0 ~&~ \tau_2
\end{pmatrix}\,,\quad
\begin{pmatrix}
\tau_1 ~&~ \tau_3\\
\tau_3 ~&~ \tau_1
\end{pmatrix}\,.
\end{equation}
Notice that there are only two independent regions of 2-complex dimension in the fundamental domain of $Sp(4,\mathbb{Z})$,
$\begin{pmatrix}
\tau_1 ~&~ 0\\
0~&~ \tau_2
\end{pmatrix}
$ and
$
\begin{pmatrix}
\tau_1 ~&~ \tau_3\\
\tau_3~&~ \tau_1
\end{pmatrix}$~\cite{Ding:2020zxw}. There are five inequivalent regions of filling 1-complex dimension in the fundamental domain of $Sp(4,\mathbb{Z})$, they are
$\begin{pmatrix}
i ~&~ 0\\
0 ~&~ \tau_2
\end{pmatrix}$,
$\begin{pmatrix}
\omega ~&~ 0\\
0 ~&~ \tau_2
\end{pmatrix}
$, $\begin{pmatrix}
\tau_1 ~&~ 0\\
0 ~&~ \tau_1
\end{pmatrix}$,
$\begin{pmatrix}
\tau_1 ~&~ 1/2\\
1/2  ~&~ \tau_1
\end{pmatrix}
$ and
$\begin{pmatrix}
\tau_1 ~&~ \tau_1/2\\
\tau_1/2 ~&~ \tau_1
\end{pmatrix}$ where $\omega\equiv e^{\frac{2\pi i}{3}}$~\cite{Ding:2020zxw}. We only consider a 1–complex dim case
$
\begin{pmatrix}
\tau_1 ~&~ \tau_1/2\\
\tau_1/2~&~ \tau_1
\end{pmatrix}
$ since in this setting the off-diagonal entries could play an important role in modular invariant potential, the remaining four cases can be obtained by considering specific values of $\tau_1$, $\tau_2$ and $\tau_3$ in the two dimensional subspaces. The six zero-dimensional inequivalent fixed points of $Sp(4,\mathbb{Z})$ are
$\begin{pmatrix}
\zeta ~&~ \zeta+\zeta^{-2}\\
\zeta+\zeta^{-2} ~&~ -\zeta^{-1}
\end{pmatrix}$, $
\begin{pmatrix}
\eta ~&~ \frac{1}{2}(\eta-1) \\
\frac{1}{2}(\eta-1) ~&~ \eta
\end{pmatrix}
$, $\begin{pmatrix}
i ~&~ 0\\
0 ~&~ i
\end{pmatrix}$,
$\begin{pmatrix}
\omega ~&~ 0\\
0 ~&~ \omega
\end{pmatrix}$,
$
\frac{i}{\sqrt{3}}\begin{pmatrix}
2 ~&~ 1\\
1 ~&~ 2
\end{pmatrix}
$,
$
\begin{pmatrix}
\omega ~&~ 0\\
0  ~&~ i
\end{pmatrix}
$ with $\zeta \equiv e^{2\pi i/5}$ and $\eta \equiv \frac{1}{3}(1 + i2\sqrt{2})$~\cite{Ding:2020zxw}.

\subsection{Absolute invariants of Siegel modular group for $g=2$}

It is known that there is a unique weight zero holomorphic function of $\tau$ away from the cusp $i\infty$ for $g=1$, which is Klein $j$-invariant. The $q$-expansion of $j$-invariant is given by
\begin{equation}\label{eq:kleinJ}
j(\tau) = \frac{1}{q} + 744 + 196884 q + 21493760 q^2 + 864299970q^3 + 20245856256q^4 + \ldots\,,
\end{equation}
where $q=e^{2\pi i\tau}$. Obviously, $j$ has a simple pole at the cusp $i\infty$, making it suitable for constructing $\alpha$-attractor models~\cite{Kallosh:2024ymt}. In the case of $g=2$, there exist three independent absolute invariants~\cite{6c36a4a1-df00-35e4-8934-01b06a1026c9}
\begin{equation}\label{eq:y123def}
y_1 = \frac{\psi_4^3}{\chi_{12}}  \,,\quad  y_2 = \frac{\psi_6^2}{\chi_{12}} \,,\quad y_3  = \psi_4^2 \psi_6 \frac{\chi_{10}}{\chi_{12}^2}\,,
\end{equation}
where $\psi_4$, $\psi_6$ are weight $4$ and weight $6$ the Eisenstein series of $Sp(4,\mathbb{Z})$, respectively, and $\chi_{10}$ and $\chi_{12}$ are cusp forms of weight $10$ and $12$, respectively. They can be constructed in terms of Siegel theta constant $\theta_{a_1a_2b_1b_2}$ with characteristic $a=(a_1,a_2)^t$, $b=(b_1,b_2)^t$ which is defined as~\cite{igusa1964}
\begin{equation}
\label{eq:thetafunc-def}
\theta_{a_1a_2b_1b_2}(\tau)= \sum_{p\in \mathbb{Z}^2} \exp\left[\pi i \left( \left(p+\frac{a}{2}\right)^t\tau \left(p+\frac{a}{2}\right) + \left(p+\frac{a}{2}\right)^t b\right)\right]\,,
\end{equation}
where $a_i,b_i\in \{0,1\}$ and $p=(p_1,p_2)^t$ is a column vector with $p_i\in \mathbb{Z}$. There are ten even characteristics modulo $2$, namely $(0000)$, $(0001)$, $(0010)$, $(0011)$, $(0100)$, $(0110)$, $(1000)$, $(1001)$, $(1100)$, and $(1111)$.
The theta constants vanish, if $ab^t$ is odd.
The Eisentein Series $\psi_4$, $\psi_6$ and cusp form $\chi_{10}$, $\chi_{12}$ in $Sp(4,\mathbb{Z})$ are given by~\cite{Igusa1967ModularFA,AOKI},
\begin{equation}\label{eq:ringX}
\begin{split}
\psi_4 &= 4 X_1^2 - 3 X_2 + 12288 X_3\,,\\
\psi_6 &= -8X_1^3 + 9 X_1 X_2 + 73728 X_1 X_3 - 27648 X_4\,,\\
\chi_{10} &=- X_2 X_4/4\,,\\
\chi_{12} &= (3X_2^2 X_3 - 2X_1 X_2 X_4 + 3072 X_4^2)/12\,,
\end{split}
\end{equation}
where $X_1$, $X_2$, $X_3$, and $X_4$ denote Siegel modular forms of respective weights $2$, $4$, $4$, and $6$ under level $2$ congruence subgroup of $Sp(4, \mathbb{Z})$. They can be expressed in terms of theta constants
\begin{equation}
\begin{split}
X_1 &= \left(\theta_{0000}^4 + \theta_{0001}^4 + \theta_{0010}^4 + \theta_{0011}^4\right)/4\,,\\
X_2 &= \left(\theta_{0000}\theta_{0001} \theta_{0010}\theta_{0011}\right)^2\,,\\
X_3 &= \left(\theta_{0100}^4 - \theta_{0110}^4\right)^2/16384\,,\\
X_4 &= \left(\theta_{0100}  \theta_{0110}\theta_{1000}\theta_{1001}\theta_{1100}\theta_{1111} \right)^2 /4096\,.
\end{split}
\end{equation}
From the expression of Siegel theta constant in Eq.~\eqref{eq:thetafunc-def}, we can determine the $q$-expansions of $\psi_4$, $\psi_6$, $\chi_{10}$ and $\chi_{12}$ as follow~\cite{ranestad20081,cmgtwo}
\begin{small}
\begin{eqnarray}
\nonumber \psi_4 &=& 1 + 240(q_1+q_2) + 2160(q_1^2+q_2^2)\,,\\
&&+ (240q_3^{-2} + 13440q_3^{-1} + 30240  + 13440q_3 + 240q_3^{2} )q_1 q_2 + \ldots\,,\\
\nonumber \psi_6 &=& 1 - 504(q_1+q_2) - 16632(q_1^2+q_2^2)\\
&&- (504q_3^{-2} -44352 q_3^{-1} -166320 -44352q_3 + 504q_3^{2} )q_1 q_2 +\ldots\,,\\
\nonumber -4\chi_{10} &=& (q_3^{-1}-2+ q_3)q_1q_2- (2q_3^{-2} + 16 q_3^{-1} -36  + 16q_3 + 2q_3^{2} )(q_1^2 q_2 + q_1 q_2^2)\\
&& - (16q_3^{-3} - 240q_3^{-2} + 240 q_3^{-1} - 32 + 240 q_3- 240q_3^{2} + 16q_3^{3} )q_1^2 q_2^2 +\ldots\,,\\
\nonumber 12\chi_{12} &=& (q_3^{-1}+ 10 + q_3)q_1q_2 +  (10 q_3^{-2} -88 q_3^{-1} -132  - 88 q_3 + 10q_3^{2} )(q_1^2 q_2 + q_1 q_2^2)\\
&& - (88q_3^{-3} - 2784q_3^{-2} + 8040 q_3^{-1} - 17600   + 8040 q_3   - 2784q_3^{2} + 88q_3^{3})q_1^2 q_2^2+\ldots\,,
\end{eqnarray}
\end{small}
where $q_j \equiv e^{2\pi i \tau_j}$. It is noted that $\chi_{10}$ vanishes when $\tau_3=0$. Then the moduli space would factorize into two tori, the absolute invariants $y_1$ and $y_2$ can be expressed in terms of the Klein $j$-invariant, while $y_3$ vanishes~\cite{6c36a4a1-df00-35e4-8934-01b06a1026c9}:
\begin{equation}\label{eq:y1y2tau30}
y_1(\tau) = j(\tau_1)j(\tau_2)\,,~~
y_2(\tau) = \big(j(\tau_1)-1728\big)\big(j(\tau_2)-1728\big)\,,~~
y_3(\tau) = 0\,,~\text{when}~\tau_3 = 0\,.
\end{equation}
The $q$-expansion of the absolute invariants $y_{1,2,3}(\tau)$ are very complicated and lengthy, consequently we don't present these cumbersome expressions here. In this work, we focus on the  modular subspaces
$\begin{pmatrix}
\tau_1 ~&~ \tau_3\\
\tau_3 ~&~ \tau_1
\end{pmatrix}$ and
$\begin{pmatrix}
\tau_1 ~&~ \tau_1/2\\
\tau_1/2 ~&~ \tau_1
\end{pmatrix}$, the $q$-expansions of the absolute invariants in these subspaces will be given in the context of specific inflationary models in the following.

\section{Siegel modular inflation models\label{sec:Model}}

In the $SL(2,\mathbb{Z})$ invariant $\alpha$-attractor models, inflationary predictions  are consistent with observations, and these models are naturally motivated by string theory. The imaginary part of the modulus can be canonically normalized, while the real part is stabilized through double exponential suppression. The Klein $j$-invariant is used to construct single-modulus inflationary models.

The Siegel modular group $Sp(2g,\mathbb{Z})$ is a natural extension of $SL(2,\mathbb{Z})$ to include multiple moduli. It can naturally arise in the compactification of string theory~\cite{Baur:2020yjl,Nilles:2021glx}. In the case $g=1$, one recovers $Sp(2,\mathbb{Z}) = SL(2,\mathbb{Z})$, which motivates us to consider the framework of modular inflation in $Sp(2g,\mathbb{Z})$. In this work, we focus on the genus $g=2$ and the number of moduli increases from one to three. The $\alpha$ parameter again appears naturally in the K\"ahler potential, as shown in Eq.~\eqref{eq:mtauKp}. For concreteness, in this work we restrict our attention to all 2–complex dim subspaces and one representative 1-complex dim subspace. Similar to the case of $SL(2,\mathbb{Z})$, the imaginary part of $\tau$ can also be canonically normalized, and the real part is  stabilized through double-exponential suppression.

In this section, we consider the Siegel modular invariant inflation in which the moduli $\tau$ play the role of inflaton fields. The Lagrangian for modular invariant inflation takes the form~\cite{Wess:1992cp}
\begin{equation}\label{eq:mod}
\frac{{\cal L}}{\sqrt{-g}} = g^{\mu\nu}{\cal K}_{\tau_{i}\bar{\tau}_j} \partial_\mu \tau_i ~\partial_\nu \bar{\tau}_j - V(\tau_i,\bar{\tau}_j)\,,
\end{equation}
where ${\cal K}_{\tau_{i}\bar{\tau}_j} = \partial_{\tau_{i}}\partial_{\bar{\tau}_j}{\cal K}$ denotes the K\"ahler metric in the moduli space, while the second term $V(\tau_j,\bar{\tau}_j)$ represents the scalar potential, which is required to be a Siegel modular invariant function of the moduli.
We substitute $\tau_i = \Re(\tau_i) + i\Im(\tau_i)$ into Eq.~\eqref{eq:mod}, normalize the kinetic term of $\Im(\tau_i)$ canonically, and analyze the dynamics of each component within the framework of Section~\ref{sec:Muti-frame}.

We begin by considering the extended $E$-model and $T$-model in two two-field inflationary frameworks, followed by the polynomial $\alpha$-attractor model in a single-field setting. All these models satisfy the Siegel modular invariance, as the scalar potentials are constructed by absolute invariants. The phenomenological predictions for inflation are studied under the large-field and slow-roll approximation.

\subsection{The subspace $\tau=\begin{pmatrix}
\tau_1~&~0\\
0~&~\tau_2
\end{pmatrix}$ \label{sec:tau12}}

This subspace is mapped into itself by the Siegel modular transformations $\gamma_{+}$ and $\gamma_{-}$ defined as~\cite{Ding:2020zxw}
\begin{small}
\begin{equation}
\gamma_{+} =
\begin{pmatrix}
a_1 ~&~ 0 ~&~ b_1 ~&~ 0\\
0 ~&~ a_4 ~&~ 0 ~&~b_4 \\
c_1 ~&~ 0 ~&~ d_1 ~&~ 0\\
0 ~&~ c_4 ~&~ 0 ~&~ d_4
\end{pmatrix}\,,\quad
\gamma_{-} =
\begin{pmatrix}
0 ~&~ a_4 ~&~ 0 ~&~ b_4\\
a_1 ~&~ 0 ~&~ b_1 ~&~0 \\
0 ~&~ c_4 ~&~ 0 ~&~ d_4 \\
c_1 ~&~ 0 ~&~ d_1 ~&~ 0
\end{pmatrix}\,,
\end{equation}
\end{small}
with $a_i d_i - b_i c_i = 1$ and $a_i,b_i,c_i,d_i \in \mathbb{Z}$ $(i=1,4)$. From Eq.~\eqref{eq:SiegelTrans}, we see that $\gamma_{+}$ and $\gamma_{-}$ act on $\text{diag}(\tau_1,\tau_2)$ as
\begin{equation}\label{eq:gammazf}
\gamma_{+}
\begin{pmatrix}
\tau_1 ~&~ 0\\
0~&~  \tau_2
\end{pmatrix}=
\begin{pmatrix}
\frac{a_1\tau_1 + b_1}{c_1\tau_1 + d_1} ~&~ 0\\
0 ~&~ \frac{a_4\tau_2 + b_4}{c_4\tau_2 + d_4}
\end{pmatrix}\,,\quad \gamma_{-}
\begin{pmatrix}
\tau_1 ~&~ 0\\
0~&~  \tau_2
\end{pmatrix}=
\begin{pmatrix}
\frac{a_4\tau_2 + b_4}{c_4\tau_2 + d_4} ~&~ 0\\
0 ~&~ \frac{a_1\tau_1 + b_1}{c_1\tau_1 + d_1}
\end{pmatrix}\,.
\end{equation}
Hence the transformation $\gamma_{+}$ acts as two independent $SL(2,\mathbb{Z})$ transformations on $\tau_1$ and $\tau_2$, while the transformation $\gamma_{-}$ exhibits a mirror symmetry $\tau_1\leftrightarrow \tau_2$ when $a_1 = d_1 = a_4 = d_4 = 1$ and $b_1 = c_1 = b_4 = c_4 = 0$. Thus the two tori characterized by $\tau_1$ and $\tau_2$ are not independent here, they are related by a mirror symmetry.

For $\tau_3 = 0$, one finds $y_3 = 0$, while $y_{1,2}$ are formed from the Klein–$J$ invariant, as shown in Eq.~\eqref{eq:y1y2tau30}. Therefore we focus on potential built from the absolute invariants $y_1$ and $y_2$. In the large $\Im(\tau_{1,2})$ limit, the invariants $y_{1,2}$ are dominated by $(q_1 q_2)^{-1}$, since the dominant term of $j(\tau_{1,2})$ is $q_{1,2}^{-1}$, so the $|y_{1,2}|^2$ can be approximately written as
\begin{equation}\label{eq:yapprox}
|y_{1,2}(\tau)|^2 \approx e^{4\pi \Im(\tau_1)} e^{4\pi \Im(\tau_2)}\,.
\end{equation}
Equipped with these relations, we proceed to construct a modular invariant potential of the form
\begin{equation}\label{Eq:potentialV1}
V(\tau_1,\tau_2) = V_0 \left( 1- \frac{\ln\beta^2}{\ln\left(|y_{2}(\tau)|^2+\beta^2\right)}\right)^2\,,
\end{equation}
where $\beta$ is some constant, we have chosen $y_2(\tau)$ as a representative example, then the minima of the potential are localized in $\tau_1=i$ or $\tau_2=i$ and they are Minkowski vacua, which facilitates subsequent analysis. If $y_1(\tau)$ instead of $y_2(\tau)$ is used in Eq.~\eqref{Eq:potentialV1}, the vacuum is shift to $\tau_1=e^{\frac{2\pi}{3}i}$ or $\tau_2=e^{\frac{2\pi}{3}i}$. The predictions for inflationary observables would not significantly change in the large field limit.
The parameter $\beta$ is introduced to ensure the logarithm is well-defined. It is remarkable that the potential $V(\tau_1,\tau_2)$ is invariant under the generic modular transformation $\gamma_{+}$ and $\gamma_{-}$ in Eq.~\eqref{eq:gammazf}.

\begin{figure}
\centering
\includegraphics[scale=0.55]{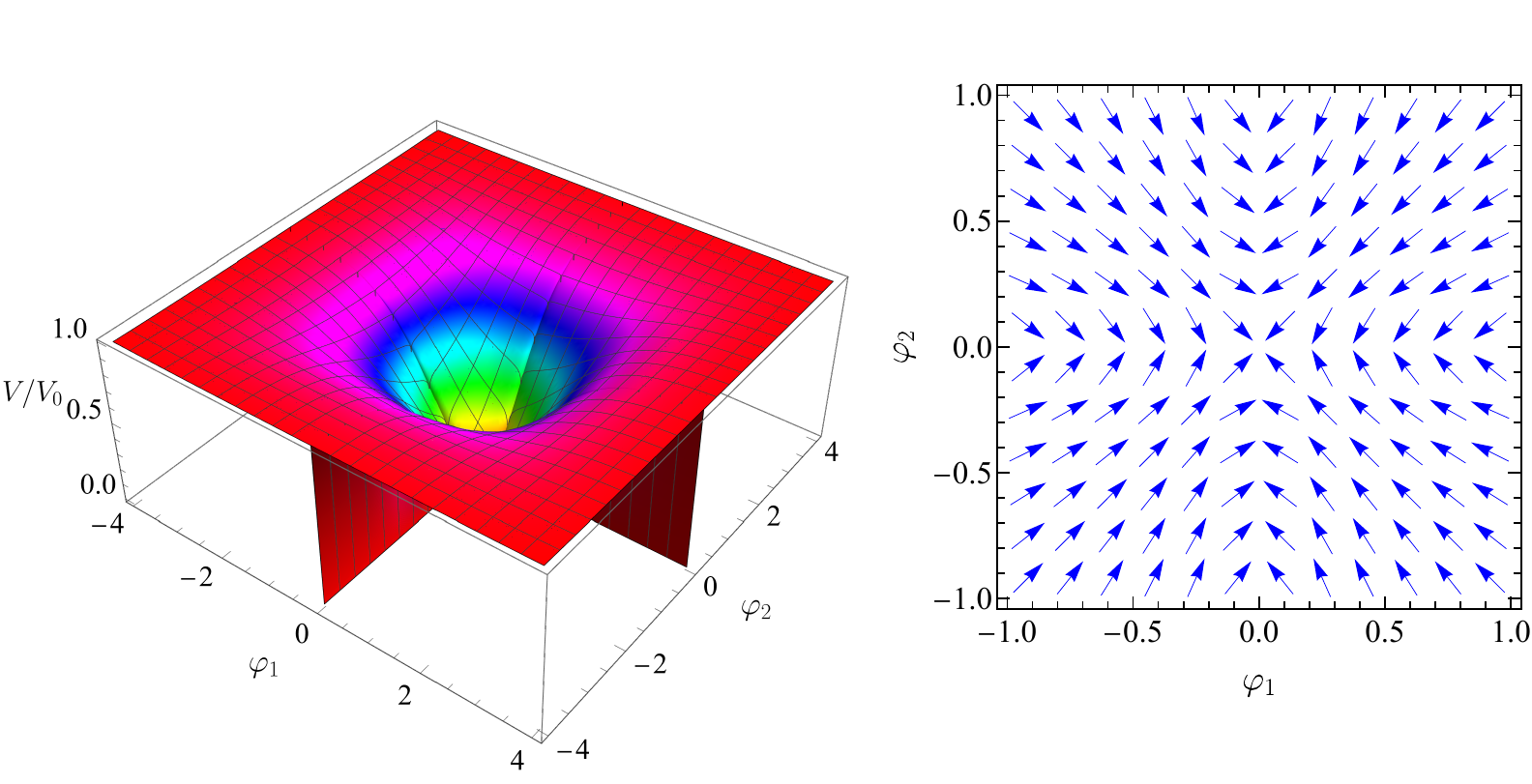}
\caption{The left panel illustrates the variation of the scalar potential $V$ in Eq.~\eqref{Eq:potentialV1}, as a function of  $\varphi_1$ and $\varphi_2$. The right panel displays the corresponding gradient vector field. Here the parameters of potential are set to $\alpha=1/3$ and $\beta=e^{4\pi}$.
}
\label{fig:plot3dreal}
\end{figure}

We now return to the discussion of the kinetic term derived from the K\"ahler potential. The Siegel K\"ahler potential in Eq.~\eqref{eq:mtauKp} reduces to the following form
\begin{equation}
\label{eq:K-diag}
\mathcal{K} \;=\; -3\alpha\,\log\!\left[-\big(\tau_1-\bar\tau_1\big)\big(\tau_2-\bar\tau_2\big)\right].
\end{equation}
Hence the kinetic terms of the complex moduli $\tau_1$ and $\tau_2$ are given by
\begin{align}
\label{eq:kinetic-tau12}
\frac{\mathcal{L}_{\rm kin}}{\sqrt{-g}}
&= \mathcal{K}_{\tau_1\bar\tau_1}\,\partial_\mu\tau_1\,\partial^\mu\bar\tau_1
+ \mathcal{K}_{\tau_2\bar\tau_2}\,\partial_\mu\tau_2\,\partial^\mu\bar\tau_2 \notag\\
&= \frac{3\alpha}{4\,\Im(\tau_1)^2}\!\left[\left(\partial_\mu\Re(\tau_1)\right)^2+\left(\partial_\mu\Im(\tau_1)\right)^2\right]
+ \frac{3\alpha}{4\,\Im(\tau_2)^2}\!\left[\left(\partial_\mu\Re(\tau_2)\right)^2+\left(\partial_\mu\Im(\tau_2)\right)^2\right]\,,
\end{align}
where $\tau_1=\Re(\tau_1)+i \Im(\tau_1)$ and $\tau_2=\Re(\tau_2)+i \Im(\tau_2)$.
It is convenient to parametrize the imaginary parts in terms of canonically normalized fields $\Im(\tau_1) =  e^{\sqrt{\tfrac{2}{3\alpha}}\, \varphi_1}$ and $
\Im(\tau_2)=  e^{\sqrt{\tfrac{2}{3\alpha}}\,\varphi_2}$. Then the kinetic term of Lagrangian takes the following form
\begin{equation}
\label{eq:Lagrangian-diag}
\frac{\mathcal{L}_{\text{kin}}}{\sqrt{-g}} =
\frac{1}{2}(\partial_\mu\varphi_1)^2 + \frac{1}{2}(\partial_\mu\varphi_2)^2
+ \frac{3\alpha}{4}\,e^{-2\sqrt{\tfrac{2}{3\alpha}}\,\varphi_1}(\partial_\mu\theta_1)^2
+ \frac{3\alpha}{4}\,e^{-2\sqrt{\tfrac{2}{3\alpha}}\,\varphi_2}(\partial_\mu\theta_2)^2
\,,
\end{equation}
where we have denoted $\theta_1 = \Re(\tau_1)$ and $\theta_2 = \Re(\tau_2)$ for simplicity of notation. In the large field limit $\varphi_1\rightarrow \infty$ and $\varphi_2\rightarrow \infty$, the Siegel modular invariant scalar potential $V(\tau_1,\tau_2)$ with $\beta=e^{4\pi}$ in Eq.~\eqref{Eq:potentialV1} can be expressed in terms of $\varphi_1$ and $\varphi_2$ as follow
\begin{equation}
V \approx V_0 \left( 1 - \frac{2}{e^{\sqrt{\frac{2}{3\alpha}}\varphi_1} +  e^{\sqrt{\frac{2}{3\alpha}}\varphi_2}} \right)^2\,.
\end{equation}
In comparison with $E$-model, the factor $e^{\sqrt{\frac{2}{3\alpha}}\varphi_1}$ is extended to $e^{\sqrt{\frac{2}{3\alpha}}\varphi_1} +  e^{\sqrt{\frac{2}{3\alpha}}\varphi_2}$~\cite{Kallosh:2015zsa}.
We plot the variation of scalar potential  in Eq.~\eqref{Eq:potentialV1} with respect to $\varphi_1$ and $\varphi_2$ in figure~\ref{fig:plot3dreal} where we have set $\theta_1=\theta_2=0$ for illustration. It can be seen that the scalar potential develops a sufficiently flat plateau in the region of large $\varphi_1$ and $\varphi_2$, which is helpful to realize the slow-roll inflation. It is clear that the degenerate vacua of the potential locate along the axes $\varphi_1=0$ and $\varphi_2=0$.
From Eq.~\eqref{eq:FeqKGf},
we can obtain the equations of motion for $\theta_1$, $\theta_2$, $\varphi_1$ and $\varphi_2$ as follow
\begin{eqnarray}
\label{eq:varphi1DY0}\ddot{\varphi}_1 + 3 H \dot{\varphi}_1 +  \sqrt{\frac{3\alpha}{2} } e^{-2\sqrt{\frac{2}{3\alpha}}\varphi_1} (\dot{\theta}_1)^2  + V_{,\varphi_1}  &=& 0\,,\\
\label{eq:varphi2DY0}\ddot{\varphi}_2 + 3 H \dot{\varphi}_2  +  \sqrt{\frac{3\alpha}{2} } e^{-2\sqrt{\frac{3\alpha}{2}}\varphi_2} (\dot{\theta}_2)^2 + V_{,\varphi_2}  &=& 0 \,,\\
\label{eq:theta1DY0}\ddot{\theta}_1 + 3 H \dot{\theta}_1 - 2 \sqrt{\frac{2}{3\alpha}} \dot{\varphi}_1\dot{\theta}_1 + \frac{2}{3\alpha} e^{2\sqrt{\frac{2}{3\alpha}}\varphi_1}V_{,\theta_1} &=& 0\,,\\
\label{eq:theta2DY0}\ddot{\theta}_2 + 3 H \dot{\theta}_2 - 2 \sqrt{\frac{2}{3\alpha}} \dot{\varphi}_2\dot{\theta}_2 + \frac{2}{3\alpha}e^{2\sqrt{\frac{2}{3\alpha}}\varphi_2}V_{,\theta_2} &=& 0\,.
\end{eqnarray}
In the slow-roll approximation and $\varphi_{1,2}\gg \sqrt{\alpha}$, the equations of $\varphi_1$ and $\varphi_2$ reduce to~\footnote{As shown in this section, the $\theta_1$ and $\theta_2$ are approximately frozen during the inflation so that $\dot{\theta}_1 \approx \dot{\theta}_2\approx 0 $.}
\begin{equation}
3 H \dot{\varphi}_1 = - V_{,\varphi_1} \,,\quad
3 H \dot{\varphi}_2 = - V_{,\varphi_2} \,,
\end{equation}
which take the standard slow-roll form. The last terms in Eqs.~\eqref{eq:theta1DY0} and \eqref{eq:theta2DY0} for the axionic fields $\theta_{1,2}$ seem large due to the presence of the exponential factor $\exp\left(2\sqrt{\tfrac{2}{3\alpha}}\varphi_{1,2}\right)$. However, we will show that $V_{,\theta_1}$ and $V_{,\theta_2}$ are in fact doubly exponentially suppressed, allowing these terms to be neglected. Consequently, under the slow-roll approximation, the $\theta_{1,2}$ fields rapidly stabilize and remain effectively frozen during inflation. In large field approximation, $y_2$ can be written as
\begin{equation}
y_2  \approx \frac{1}{q_1q_2}\left[ 1 - 984 q_1 - 984  q_2\right]\,,
\end{equation}
where we have neglected all higher-order terms. Thus the expression of $|y_2|^2$ in large field limit is given by
\begin{small}
\begin{eqnarray}
\nonumber |y_2|^2 &\approx& |q_1q_2|^{-2}
\left[ 1 -984 (q_1 + \bar{q}_1) -984 (q_2 + \bar{q}_2) \right]\\
&\approx& e^{4\pi \left(e^{\sqrt{\frac{2}{3\alpha}}\varphi_1} + e^{\sqrt{\frac{2}{3\alpha}}\varphi_2}\right)}
\left[ 1 -1968 e^{-2\pi   e^{\sqrt{\frac{2}{3\alpha}}\varphi_1}}\cos(2\pi  \theta_1) -1968 e^{-2\pi   e^{\sqrt{\frac{2}{3\alpha}}\varphi_2}}\cos(2\pi  \theta_2)\right]\,.~~~~~~
\end{eqnarray}
\end{small}
Straightforward calculations show that the $\frac{2}{3\alpha} e^{2\sqrt{\tfrac{2}{3\alpha}}\varphi_1}V_{,\theta_1}$ and $\frac{2}{3\alpha} e^{2\sqrt{\tfrac{2}{3\alpha}}\varphi_2}V_{,\theta_2}$ are doubly exponentially suppressed and can be safely neglected in equation of motion. Hence the equations for the evolution of $\theta_1$ and $\theta_2$ become
\begin{equation}
\ddot{\theta}_1 + 3 H \dot{\theta}_1 \left(1 - \frac{2}{3} \sqrt{\frac{2}{3\alpha}} \frac{\dot{\varphi}_1}{H} \right)= 0\,,\quad  \ddot{\theta}_2 + 3 H \dot{\theta}_2 \left(1 - \frac{2}{3} \sqrt{\frac{2}{3\alpha}} \frac{\dot{\varphi}_2}{H} \right)= 0
\,,
\end{equation}
where the terms $\dot{\varphi}_1/H$ and $\dot{\varphi}_2/H$ can be omitted in accordance with the slow-roll approximation\footnote{From Eqs.~\eqref{eq:approx} and \eqref{eq:phiadotapprox}, we have
\begin{equation}\nonumber
\frac{\dot{\varphi}_1}{H} \approx - \frac{V_{\varphi_1}}{V}  \,,\quad \frac{\dot{\varphi}_2}{H} \approx - \frac{V_{\varphi_2}}{V} \,.
\end{equation}
In our concerned case, the slow-roll parameter in Eq.~\eqref{eq:epVoprep} becomes
\begin{equation}\nonumber
\epsilon_V = \frac{1}{2}\frac{V_{,\varphi_1}^2 + V_{,\varphi_2}^2}{V^2}\,,
\end{equation}
It then follows that
\begin{equation}\nonumber
\sqrt{2\epsilon_V} > \left|\frac{V_{,\varphi_1}}{V}\right|\,,\quad \sqrt{2\epsilon_V} > \left|\frac{V_{,\varphi_2}}{V}\right|\,,
\end{equation}
which implies
\begin{equation}\nonumber
\left|\frac{\dot{\varphi}_1}{H}\right| < \sqrt{2\epsilon_V} \ll 1\,, \qquad
\left|\frac{\dot{\varphi}_2}{H}\right| < \sqrt{2\epsilon_V} \ll 1\,.
\end{equation}
}, we have
\begin{equation}\label{eq:theta1theta2}
\ddot{\theta}_1 + 3 H \dot{\theta}_1 = 0\,,\quad \ddot{\theta}_2 + 3 H \dot{\theta}_2 = 0\,.
\end{equation}
The solution of equations are~\cite{Kallosh:2024kgt}
\begin{equation}
\theta_1 =\theta_{1}^{(0)} \left(1-e^{-3Ht}\right)\,,\quad \theta_2 = \theta_{2}^{(0)} \left(1-e^{-3Ht}\right)\,,
\end{equation}
where $\theta_{1}^{(0)}$ and $\theta_{2}^{(0)}$ are constants. It is noticed that $H$ is approximately constant during inflation. Consequently, the fields $\theta_{1,2}$ freeze to $\theta_{1,2}^{(0)}$ within a few $e$-foldings. For simplicity, we set $\theta_{1,2}^{(0)} = 0$. Let us consider the following three particular Siegel modular transformations
\begin{eqnarray}
\nonumber \gamma_1&=&{\cal T}_1 {\cal S} {\cal T}_1 {\cal S}^{-1} {\cal T}_1 =
\begin{pmatrix}
0 ~&~ 0 ~&~ 1 ~&~ 0\\
0 ~&~ 1 ~&~ 0 ~&~0 \\
-1 ~&~ 0 ~&~ 0 ~&~ 0\\
0 ~&~ 0 ~&~ 0 ~&~ 1
\end{pmatrix}\,,\\
\nonumber \gamma_2 &=& {\cal T}_2 {\cal S} {\cal T}_2 {\cal S}^{-1} {\cal T}_2 =
\begin{pmatrix}
1 ~&~ 0 ~&~ 0 ~&~ 0\\
0 ~&~ 0 ~&~ 0 ~&~1 \\
0 ~&~ 0 ~&~ 1 ~&~ 0\\
0 ~&~ -1 ~&~ 0 ~&~ 0
\end{pmatrix}\,,\\
\gamma_3 &=& \left({\cal S} {\cal T}_3\right)^3 =
\begin{pmatrix}
0 ~&~ 1 ~&~ 0 ~&~ 0\\
1 ~&~ 0 ~&~ 0 ~&~0 \\
0 ~&~ 0 ~&~ 0 ~&~ 1\\
0 ~&~ 0 ~&~ 1~&~ 0
\end{pmatrix}\,.
\end{eqnarray}
Their action on the moduli $\tau_1$ and $\tau_2$ reads as
\begin{eqnarray}
\hskip-0.1in \gamma_{1}
\begin{pmatrix}
\tau_1 ~& 0\\
0~&  \tau_2
\end{pmatrix}=
\begin{pmatrix}
-\frac{1}{\tau_1} ~& 0\\
0 ~& \tau_2
\end{pmatrix}
\,,~~
\gamma_{2}
\begin{pmatrix}
\tau_1 ~& 0\\
0~&  \tau_2
\end{pmatrix}=
\begin{pmatrix}
\tau_1 ~& 0\\
0 ~& -\frac{1}{\tau_2}
\end{pmatrix}\,,~~ \gamma_{3}
\begin{pmatrix}
\tau_1 ~& 0\\
0~&  \tau_2
\end{pmatrix}=
\begin{pmatrix}
\tau_2 ~& 0\\
0 ~& \tau_1
\end{pmatrix}\,.
\end{eqnarray}
Therefore the modular transformations of the canonically normalized fields $\varphi_1$ and $\varphi_2$ are given by
\begin{eqnarray}
\nonumber \gamma_1&:& \varphi_1\rightarrow -\varphi_1, ~\varphi_2\rightarrow \varphi_2\,,\\
\nonumber \gamma_2&:& \varphi_1\rightarrow \varphi_1, ~\varphi_2\rightarrow -\varphi_2\,,\\
 \gamma_3&:& \varphi_1\leftrightarrow \varphi_2\,,
\end{eqnarray}
which leave the scalar potential invariant.

\begin{figure}
\centering
\includegraphics[scale=0.6]{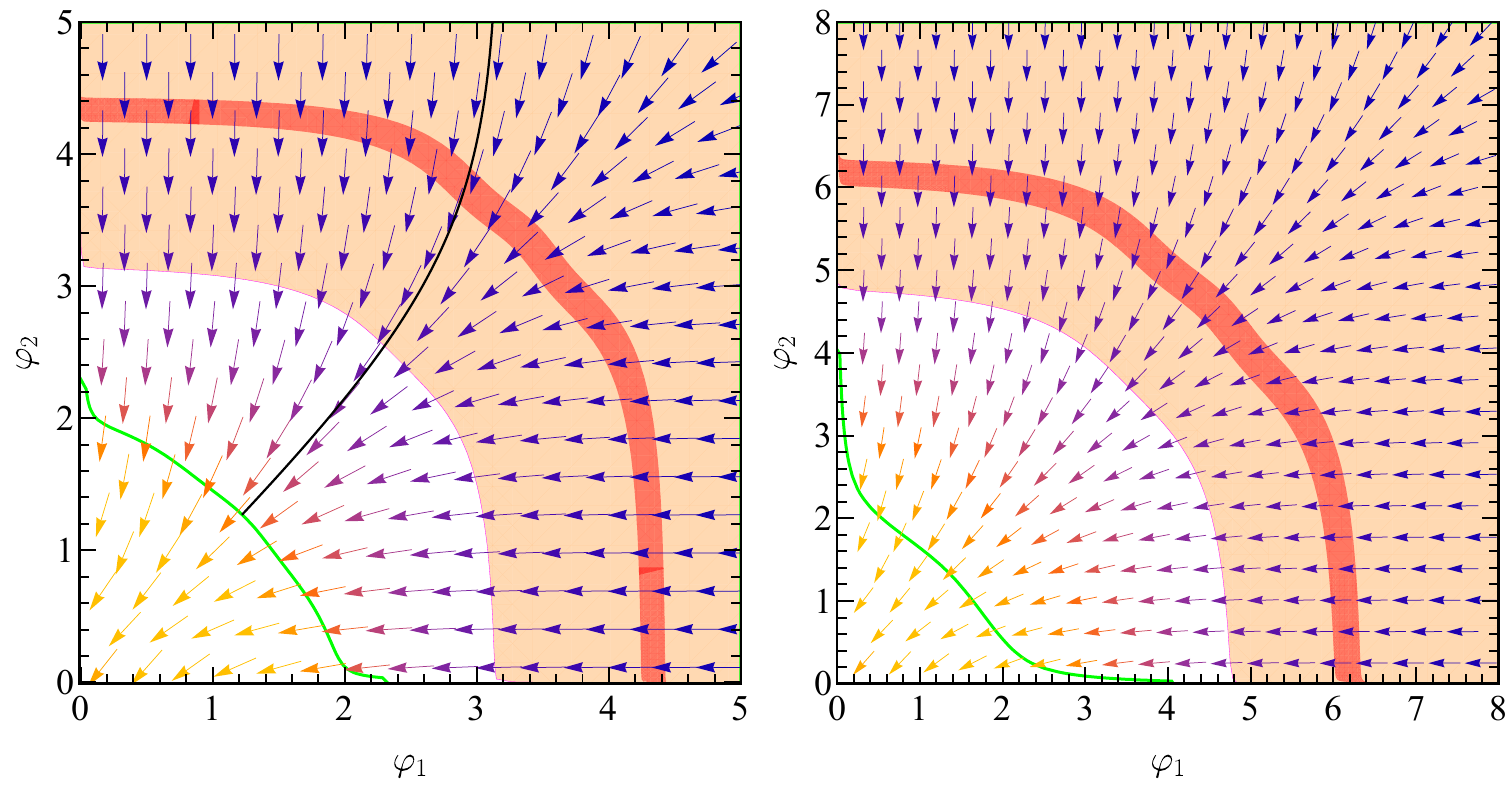}
\caption{Vector plots of the potential gradients for the potential in Eq.~\eqref{Eq:potentialV1}. The left and right panels correspond to $\alpha=1/3$ and $\alpha=1$, respectively, both with $\beta=e^{4\pi}$. The red region denotes the spectral index range $0.9602 < n_s < 0.9693$~\cite{Planck:2018vyg}. The orange region shows where the tensor-to-scalar ratio satisfies $r \leq 0.036$~\cite{BICEP:2021xfz}, with its boundary indicated by a thin pink line. The green line represents the condition $|\eta_H| = 1$, marking the end of inflation.}
\label{fig:inflationmodel2}
\end{figure}

\begin{figure}
\centering
\includegraphics[scale=0.6]{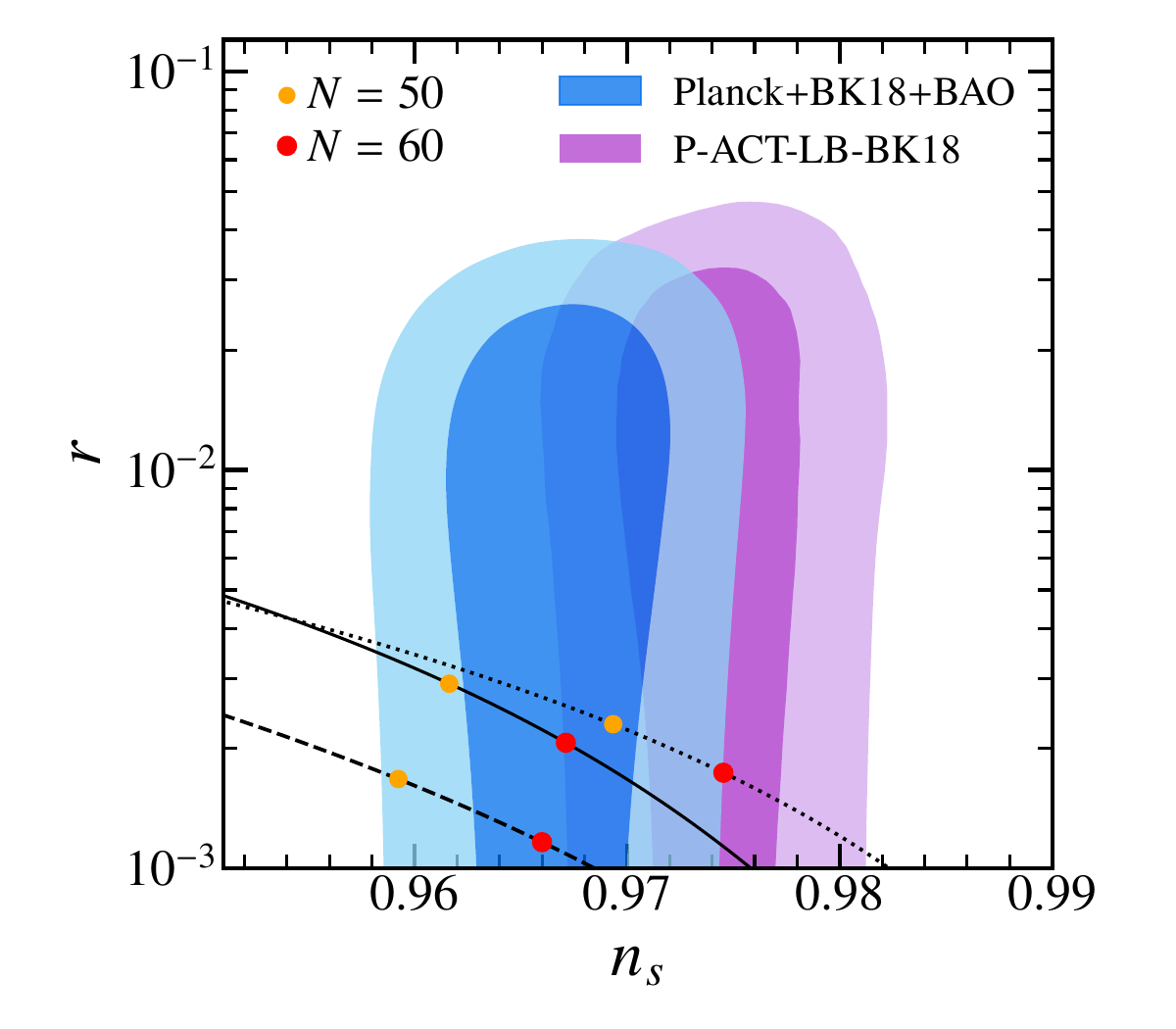}
\caption{The solid, dashed, and dotted lines represent the theoretical $(n_s, r)$ predictions from the models in Eqs.~\eqref{Eq:potentialV1}, \eqref{eq:tau13potential}, and \eqref{eq:vact}, with $\alpha = 1/3$ and distinct initial conditions , respectively. The blue contours denote the $68\%$ and $95\%$ confidence regions obtained from the combined Planck 2018, BICEP/Keck 2018, and BAO data~\cite{BICEP:2021xfz}, whereas the purple contours represent the results from the combined Planck 2018, ACT, BAO, BICEP/Keck 2018~\cite{ACT:2025tim}. The small orange and large red dots indicate the predictions for $N = 50$ and $N = 60$, respectively.}
\label{fig:inflationmodelwhole}
\end{figure}

The CMB observables, like $r,n_s,N$ are evaluated at the moment of horizon crossing. We use $(\varphi_{1*},\varphi_{2*})$ to denote the corresponding field values. In figure~\ref{fig:inflationmodel2}, we show the constraints on $(\varphi_{1*},\varphi_{2*})$ from the measured tensor to scalar ratio $r$ and spectral index $n_s$ by Planck. The orange region shows the field space where the tensor to scalar ratio $r<0.036$ and the red region shows the field space where the spectral index $n_s$ takes the value between $0.9602<n_s<0.9693$. Moreover, the green line shows the boundary where the slow-roll condition fails and inflation terminates. Inflation compatible with CMB observations requires an overlap between regions allowed by $r$ and $n_s$. In figure~\ref{fig:inflationmodelwhole}, we show the prediction of the model by varying the CMB field values along a specific gradient flow shown in figure~\ref{fig:inflationmodel2}. Along the solid line in figure~\ref{fig:inflationmodelwhole}, smaller e-folding corresponds to smaller values of $n_s$ and larger values of $r$.
For $50$ and $60$ e-folds, we have
\begin{equation}
\begin{aligned}
& N=50\,,\quad\varphi_{1*}=2.8726\,,\quad \varphi_{2*}=3.6694\,,\quad  n_s=0.9616\,,\quad r = 0.0029 \,,\\
&N=60\,,\quad \varphi_{1*}=2.9211\,,\quad \varphi_{2*}=3.8323\,,\quad   n_s=0.9671\,,\quad r =0.0021 \,.
\end{aligned}
\end{equation}
\begin{table}
\begin{center}
\begin{tabular}{|| c c c c c c c||}
\hline
$\varphi_{1*}$ & $\varphi_{2*}$ & $n_s$ & $r$ & $\alpha_s$ & $N$ & $f_{\text{NL}}$ \\ [0.5ex]
\hline\hline
0.2468 & 4.3660 & 0.9667 & $1.1\times 10^{-3}$& $-5.5\times 10^{-4}$ &$58.75$ & $-0.0138$\\
\hline
0.4042 & 4.2840 & 0.9627 & $1.4\times 10^{-3}$ &$-6.9\times 10^{-4}$ & $52.61$   & $-0.0155$\\
\hline
2.6081 & 3.9132 & 0.9619 & $2.3\times 10^{-3}$ & $-6.0\times 10^{-4}$ & $54.50$  & $-0.0158$\\
\hline
3.1200 & 3.6045 & 0.9659 & $2.4\times 10^{-3}$ & $-5.9\times 10^{-4}$ & $55.11$ & $-0.0141$ \\
\hline
4.2191 & 1.7223 & 0.9628 &$1.5\times 10^{-3}$ & $-6.6\times 10^{-4}$ & $56.83$ & $-0.0154$\\ [1ex]
\hline
\end{tabular}
\end{center}
\caption{The predictions for different sample points and inflationary trajectories for potential~\eqref{Eq:potentialV1} with $\alpha=1/3$ and $\beta=e^{4\pi}$.}
\label{tab:traj2}
\end{table}
Several typical choices of CMB field values starting points
and their corresponding predictions for inflation observables are shown in table~\ref{tab:traj2}. We see that the e-folds in the region $50$–$60$ can be produced, and the resulting $n_s$ lies within the observationally allowed range. The tensor-to-scalar ratio $r$ is of order ${\cal O}(10^{-3})$ and the running of the spectral index $\alpha_s$ is of order $-{\cal O}(10^{-4})$, which are within the sensitivity of future experiments~\cite{LiteBIRD:2022cnt,Abazajian:2019eic}. Moreover, we can also see the predicted non-Gaussianity is of order ${\cal O}(10^{-2})$, consistent with current observational constraints~\cite{Planck:2019kim}.

\subsection{The subspace $\tau=\begin{pmatrix}
\tau_1~&~\tau_3\\
\tau_3~&~\tau_1
\end{pmatrix}$\label{sec:tau13}}

In this case, the presence of off-diagonal entries in the moduli $\tau$ makes the canonical normalization of its imaginary components more challenging than the diagonal case $\tau=\mathrm{diag}(\tau_1,\tau_2)$. Nevertheless, with a suitable parametrization, the imaginary parts of $\tau$ can still be normalized. The K\"ahler potential in this case is taken to be
\begin{equation}
{\cal K}=- 3\alpha \ln \left[(\tau_3 - \bar{\tau}_3)^2-(\tau_1 - \bar{\tau}_1)^2\right]\,.
\end{equation}
Consequently the kinetic terms are of the following form:
\begin{equation}
\frac{{\cal L}_{\text{kin}}}{\sqrt{-g}} =  \frac{3\alpha}{4}\frac{1}{\Im(z_1)^2} \partial_\mu z_1\partial^\mu\overline{z}_1 +  \frac{3\alpha}{4}\frac{1}{\Im(z_2)^2} \partial_\mu z_2\partial^\mu\overline{z}_2\,,
\end{equation}
where
\begin{equation}\label{eq:z1z2}
z_1 = \frac{\tau_3 - \tau_1 }{\sqrt{2}}\,,\quad z_2 = \frac{\tau_3 + \tau_1 }{\sqrt{2}}\,.
\end{equation}
The kinetic terms of imaginary parts can be canonically normalized by field redefinition as follows
\begin{equation}\label{eq:Imz1z2}
\Im(z_1) = -\frac{1}{\sqrt{2}}e^{\sqrt{\tfrac{2}{3\alpha}}\, \varphi_1}\,,\quad
\Im(z_2)= \frac{1}{\sqrt{2}} e^{\sqrt{\tfrac{2}{3\alpha}}\,\varphi_2}\,.
\end{equation}
Then the kinetic term of Lagrangian takes the following form
\begin{equation}
\frac{{\cal L}_{\text{kin}}}{\sqrt{-g}} = \frac{3\alpha}{2}e^{-2\sqrt{\frac{2}{3\alpha}} \varphi_1 }(\partial_\mu \theta_1 )^2 +
\frac{3\alpha}{2}e^{-2\sqrt{\frac{2}{3\alpha}} \varphi_2 }(\partial_\mu \theta_2 )^2+  \frac{1}{2}(\partial_\mu \varphi_2)^2 + \frac{1}{2}(\partial_\mu \varphi_1)^2\,,
\end{equation}
where we have denoted $\theta_1 = \Re(z_1)$ and $\theta_2 = \Re(z_2)$.
Consequently, the equations of motion for $\theta_1$, $\theta_2$, $\varphi_1$ and $\varphi_2$ are given by
\begin{eqnarray}
\label{eq:theta1DY}\ddot{\theta}_1 + 3 H \dot{\theta}_1 - 2 \sqrt{\frac{2}{3\alpha}} \dot{\varphi}_1\dot{\theta}_1  + \frac{1}{3\alpha}e^{2\sqrt{\frac{2}{3\alpha}}\varphi_1 }V_{,\theta_1} &=& 0\,,\\
\label{eq:theta2DY}\ddot{\theta}_2 + 3 H \dot{\theta}_2 - 2 \sqrt{\frac{2}{3\alpha}} \dot{\varphi}_2\dot{\theta}_2 + \frac{1}{3\alpha} e^{2\sqrt{\frac{2}{3\alpha}} \varphi_2 }V_{,\theta_2} &=& 0\,,\\
\label{eq:varphi1DY}\ddot{\varphi}_1 + 3 H \dot{\varphi}_1 +  \sqrt{6\alpha}\, e^{-2\sqrt{\frac{2}{3\alpha}} \varphi_1 }\dot{\theta}_1^2 + V_{,\varphi_1}  &=& 0\,,\\
\label{eq:varphi2DY}\ddot{\varphi}_2 + 3 H \dot{\varphi}_2   +  \sqrt{6\alpha}\, e^{-2\sqrt{\frac{2}{3\alpha}} \varphi_2 } \dot{\theta}_2^2 + V_{,\varphi_2}  &=& 0 \,.
\end{eqnarray}
The fields $\varphi_{1,2}$, corresponding to the imaginary parts of the moduli, act as the inflaton candidates. Their equations of motion are coupled to the dynamics of $\theta_{1,2}$, whose evolution must be analyzed using the explicit form of the scalar potential.

We next construct a modular invariant inflationary potential within this subspace. The scalar potential is constructed from absolute invariants, which can be expanded in terms of $q_1$ and $q_3$ as follows:
\begin{eqnarray}
y_1 &=& \frac{q_3}{q_1^2}~\frac{1 + 1440 q_1 + \left(\frac{720}{q_3^2} + \frac{40320}{q_3} + 794880 + 40320 q_3 + 720 q_3^2\right)q_1^2 + {\cal O}(q_1^3)}{\frac{1}{12} + \frac{5}{6}q_3 + \frac{1}{12}q_3^2 + \frac{1}{3}\left( 5- 44 q_3 - 66q_3 ^2 -44 q_3^3 + 5q_3^4\right)\frac{q_1}{q_3}+ {\cal O}(\frac{q_1^2}{q_3^2})}\,,\\
y_2 &=& \frac{q_3}{q_1^2}~\frac{1 - 2016 q_1 - \left(\frac{1008}{q_3^2} - \frac{88704}{q_3} - 1282176 - 88704 q_3 + 1008 q_3^2\right)q_1^2 + {\cal O}(q_1^3)}{\frac{1}{12} + \frac{5}{6}q_3 + \frac{1}{12}q_3^2 + \frac{1}{3}\left( 5- 44 q_3 - 66q_3 ^2 -44 q_3^3 + 5q_3^4\right)\frac{q_1}{q_3}+ {\cal O}(\frac{q_1^2}{q_3^2})}\,,\\
y_3 &=& \frac{q_3}{q_1^2}~\frac{(-\frac{1}{4} + \frac{1}{2}q_3 -\frac{1}{4}q_3^2) + (1+ 20q_3 -42 q_3^2 + 20q_3^3 + q_3^4)\frac{q_1}{q_3}+ {\cal O}(\frac{q_1^2}{q_3^2})}{ \frac{1}{144}(1+ 20 q_3 + 102 q_3^2 + 20 q_3^3 + q_3^4)+ {\cal O}(\frac{q_1}{q_3})}\,,
\end{eqnarray}
where $|q_1|$ and $|q_1/q_3|$ are small in the large-field limit
\begin{equation}
\begin{split}
\left|q_1 \right|
&= \exp\left[-  \pi \left( e^{\sqrt{\frac{2}{3\alpha}} \varphi_1 } + e^{\sqrt{\frac{2}{3\alpha}} \varphi_2 }\right) \right]\,, \\
\left|q_3\right| &= \exp\left[-\pi (e^{\sqrt{\tfrac{2}{3\alpha}}\varphi_2} - e^{\sqrt{\tfrac{2}{3\alpha}}\varphi_1})\right] \,,\quad \\
\left|\frac{q_1}{q_3}\right| & = \exp\left[-2 \pi e^{\sqrt{\frac{2}{3\alpha}}\varphi_1} \right]\,.
\end{split}
\end{equation}
In this work, we focus on the region $\varphi_2 > \varphi_1$ which implies $|q_3| < 1$.
Thus we see that all three invariants $y_{1,2,3}(\tau)$ are dominated by the term $q_3/q_1^2$. In large field approximation, $\log |y_{1,2,3}|^2$ can be expressed as
\begin{equation}
\frac{1}{8\pi}\log|y_{1,2,3}|^2 \approx \frac{3e^{\sqrt{\frac{2}{3\alpha}}\varphi_1} +  e^{\sqrt{\frac{2}{3\alpha}}\varphi_2}}{4}\,.
\end{equation}
We are ready to construct modular invariant potential as follow
\begin{equation}\label{eq:tau13potential}
V= V_0 \left(\frac{\ln(|y_3|^2+\beta^2)-\ln\beta^2}{\ln(|y_3|^2+\beta^2)+\ln\beta^2}\right)^2\,,
\end{equation}
where $\beta$ is a constant to keep logarithm remained well-defined.
\begin{figure}
\centering
\includegraphics[scale=0.55]{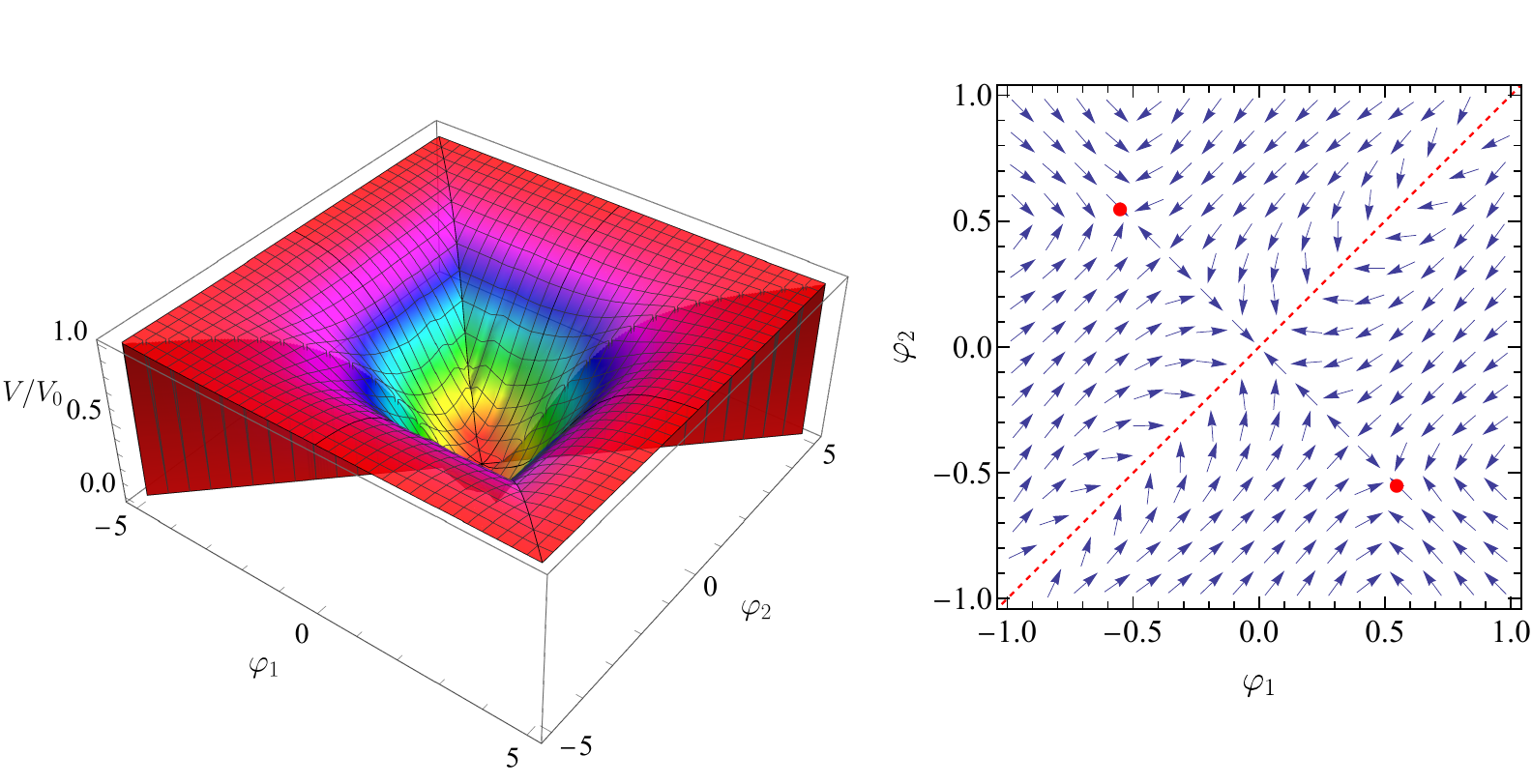}
\caption{The left panel displays the variation of the scalar potential $V$ in Eq.~\eqref{eq:tau13potential} with respect to  $\varphi_1$ and $\varphi_2$. The right panel shows the gradient vector field of this potential , where the dashed lines indicate the global minima and the dots denote the local minima. The parameters are set to $\alpha=1/3$ and $\beta=e^{4\pi}$.}
\label{fig:plot3d13}
\end{figure}
Since $y_3(\tau)=0$ when $\tau_3=0$, the scalar potential has degenerate global minima at $\tau=\text{diag}(\tau_1,\tau_1)$ for any value of the modulus $\tau_1$ with $\Im(\tau_1)>0$. The potential exactly vanishes at the minima so that the extremum is Minkowski. From Eqs.~\eqref{eq:z1z2} and \eqref{eq:Imz1z2}, we see that the moduli $\tau_3=0$ corresponds to $\varphi_1=\varphi_2$ and $\theta_1=-\theta_2$. The variation of potential $V$ with respect to $\varphi_1$ and $\varphi_2$ is displayed in figure~\ref{fig:plot3d13} where we have taken $\theta_{1,2}=0$. This potential possesses a sufficiently flat plateau to realize slow-roll inflation for large values of $\varphi_1$ and $\varphi_2$. In this limit, the potential in Eq.~\eqref{eq:tau13potential} with $\beta=e^{4\pi}$ can be expressed in terms of $\varphi_1$ and $\varphi_2$ as follow:
\begin{equation}
V\approx V_0\left(\frac{3e^{\sqrt{\frac{2}{3\alpha}}\varphi_1} +  e^{\sqrt{\frac{2}{3\alpha}}\varphi_2}-4}{3e^{\sqrt{\frac{2}{3\alpha}}\varphi_1} +  e^{\sqrt{\frac{2}{3\alpha}}\varphi_2}+4}\right)^2\,.
\end{equation}
In the limit $\varphi_2 \rightarrow 0$, the above potential reduces to the well-known $T$-model of inflation~\cite{Kallosh:2015zsa}. In large field limit, $|y_3|^2$ can be written as approximately
\begin{small}
\begin{equation}
|y_3|^2 \approx 36^2 \exp\left[2\pi\left(3e^{\sqrt{\frac{2}{3\alpha}}\varphi_1} + e^{\sqrt{\frac{2}{3\alpha}}\varphi_2}\right)\right]\left(1 -4  |q_1|\cos(2\pi\theta_2) -8\left|\frac{q_1}{q_3}\right|\cos[2\pi(\theta_2-\theta_1)]\right)\,.
\end{equation}
\end{small}
Consequently, the terms $\frac{1}{3\alpha}e^{2\sqrt{\frac{2}{3\alpha}}\varphi_1 }V_{,\theta_1}$, $\frac{1}{3\alpha}e^{2\sqrt{\frac{2}{3\alpha}}\varphi_2 }V_{,\theta_2}$ are double exponential suppressed, thus they can be neglected in Eqs.~\eqref{eq:theta1DY} and \eqref{eq:theta2DY}. The equations of motion of $\theta_1$ and $\theta_2$ then reduce to
\begin{equation}
\label{eq:thetaapprox}
\ddot{\theta}_1 + 3 H \dot{\theta}_1 - 2 \sqrt{\frac{2}{3\alpha}} \dot{\varphi}_1\dot{\theta}_1 = 0\,,\quad
\ddot{\theta}_2 + 3 H \dot{\theta}_2 - 2 \sqrt{\frac{2}{3\alpha}} \dot{\varphi}_2\dot{\theta}_1 = 0\,.
\end{equation}
Under the slow-roll approximation, the third terms in above equations are negligible. The fields $\theta_{1,2}$ would rapidly freeze into their initial values within a few $e$-folding~\cite{Kallosh:2024kgt}. To simplify the analysis, we could set $\theta_1 = \theta_2 = 0$ at the beginning of inflation. Then the Siegel modular transformation ${\cal S}$ acting on $\tau$ implies the symmetry $\varphi_1 \rightarrow -\varphi_1$ and $\varphi_2 \rightarrow -\varphi_2$. Moreover, another Siegel modular transformation ${\cal P}= (\gamma_2)^2 = \mathrm{diag}(1, -1, 1, -1)$ interchanges $\varphi_1$ and $\varphi_2$. The scalar potential is invariant under both ${\cal S}$ and ${\cal P}$ transformations, as can be seen from figure~\ref{fig:plot3d13}.

In the large-field regime, $\varphi_{1,2}\gg \sqrt{\alpha}$ with ${\theta}_{1,2}$ stabilized, the contribution of the $\dot{\theta}_1^2$ and $\dot{\theta}_2^2$ terms in Eqs.~\eqref{eq:varphi1DY} and~\eqref{eq:varphi2DY} can be neglected. Under slow-roll approximation, the evolution equations of $\varphi_1$ and $\varphi_2$ become
\begin{equation}
3H \dot{\varphi}_1   \approx -V_{,\varphi_1}\,, \quad 3H \dot{\varphi}_2   \approx -V_{,\varphi_2}\,.
\end{equation}
The dash line in figure~\ref{fig:inflationmodelwhole} shows the $(n_s,r)$ prediction of the model when choosing the parameters $\alpha=1/3$, $\beta=e^{4\pi}$ and varying the CMB field values over a specific gradient flow, originated from $\varphi_{1}=0.566$, $\varphi_{2}=5.703$. The number of e-folds between 50 and 60 is allowed by Planck observation~\cite{BICEP:2021xfz}, and the typical points for $N = 50$ and $N = 60$ are indicated by the orange and red dots, respectively
\begin{equation}
\begin{aligned}
&N=50\,,\quad \varphi_{1*}=0.5648\,,\quad \varphi_{2*}=5.1930\,,\quad   n_s=0.9592\,,\quad r = 0.0017 \,,\\
&N=60\,,\quad\varphi_{1*}=0.5654\,,\quad \varphi_{2*}=5.3230\,,\quad   n_s=0.9660\,,\quad r =0.0012 \,.
\end{aligned}
\end{equation}
Some specific values of $\left(\varphi_{1*}, \varphi_{2*}\right)$, together with their corresponding predictions for the inflationary observables, are presented in table~\ref{tab:tau13}. As shown in the table, these models predict values of $n_s$ and $\alpha_s$ that are consistent with the Planck~\cite{Planck:2018vyg}. The running $\alpha_s$ falls in the range of $-{\cal O}(10^{-3})$ to $-{\cal O}(10^{-4})$. The PICO mission is expected to measure these parameters with sensitivities of
$\sigma(n_s)<0.0015$ and $\sigma(\alpha_s)< 0.002$~\cite{NASAPICO:2019thw}.
Our models predict $r$ of order ${\cal O}(10^{-3})$, which is smaller than the current upper bound in~\cite{BICEP:2021xfz}. However, it can be tested by future experiments such as LiteBIRD~\cite{LiteBIRD:2022cnt}, AliCPT~\cite{Li:2017drr}, and SO~\cite{SimonsObservatory:2025avm}, which aim to achieve a precision of $\sigma(r) \sim 0.001$. The predicted non-Gaussianity is slightly larger, of order ${\cal O}(10^{-2})$, in agreement with current observational constraints~\cite{Planck:2019kim}. The future spectroscopic survey SPHEREx is expected to improve the constraint to $\sigma(f_{\rm NL}) = 0.87$~\cite{SPHEREx:2014bgr}. The upcoming missions such as LSST~\cite{LSSTDarkEnergyScience:2012kar}, PUMA~\cite{PUMA:2019jwd} and Euclid~\cite{Castorina:2020zhz} will also place stringent limits on $f_{\rm NL}$.
Thus our models will be ruled out if these surveys were to detect a significant non-Gaussian signal.
\begin{table}
\begin{center}
\begin{tabular}{|| c c c c c c c||}
\hline
$\varphi_{1*}$ & $\varphi_{2*}$ & $n_s$ & $r$ & $\alpha_s$ & $N$ & $f_{\text{NL}}$ \\ [0.5ex]
\hline\hline
0.5017 &  5.3154 & 0.9656 & $1.2\times 10^{-3}$ & $-5.9\times 10^{-4}$ & $58.51$  & $-0.0143$\\
\hline
0.7512 & 5.2233 & 0.9611 & $1.5\times 10^{-3}$ & $-7.5\times 10^{-4}$ & $52.32$  & $-0.0161$\\
\hline
1.4259 & 5.2221 & 0.9623 & $1.5\times 10^{-3}$ & $-6.9\times 10^{-4}$  & 56.10   & $-0.0156$ \\
\hline
2.0361 & 5.1861 & 0.9634 & $1.5\times 10^{-3}$ &  $-6.3\times 10^{-4}$ & $60.11$   & $-0.0152$\\
\hline
3.6991 & 4.0581 & 0.9627 & $2.9\times 10^{-3}$ & $-7.1\times 10^{-4}$ & $51.86$  & $-0.0154$\\
[1ex]
\hline
\end{tabular}
\end{center}
\caption{The predictions for different sample points and inflationary trajectories for the potential Eq.~\eqref{eq:tau13potential} with $\alpha=1/3$ and $\beta=e^{4\pi}$.}
\label{tab:tau13}
\end{table}

\subsection{The subspace $\tau=\begin{pmatrix}
\tau_1~&~\tau_1/2\\
\tau_1/2~&~\tau_1
\end{pmatrix}$\label{sec:tau11}}

Although this subspace contains only one modulus, its symmetry under $Sp(4,\mathbb{Z})$ is different from the modular symmetry $SL(2,\mathbb{Z})$. In this case, the K\"ahler potential is the following form
\begin{equation}
{\cal K}=-3\alpha \log\left[-\frac{3}{4}(\tau_1- \bar{\tau}_1)^2\right]\,,
\end{equation}
thus the corresponding kinetic term is given by
\begin{equation}
\frac{{\cal L}_{\text{kin}}}{\sqrt{-g}}=\frac{3\alpha}{2}\frac{\partial_\mu\tau_1\partial^\mu \bar{\tau}_1 }{\Im(\tau_1)^2} = \frac{3\alpha}{2}e^{-2\sqrt{\frac{1}{3\alpha}}\varphi}(\partial_\mu \theta)^2 + \frac{1}{2}\left(\partial_\mu\varphi\right)^2\,,
\end{equation}
where we have introduced a canonically normalized field $\varphi = \sqrt{3\alpha} \ln \Im(\tau_1)$ and the modulus is parameterized as $\tau_1 = \theta + ie^{\sqrt{\frac{1}{3\alpha}}\varphi}$.
Our goal is to construct a Siegel modular invariant potential, inspired by the polynomial $\alpha$-attractor model~\cite{Kallosh:2022feu,Bhattacharya:2022akq}.
In this case, the $q$-expansion of the three absolute invariants $y_{1,2,3}$ are found to be
\begin{eqnarray}
\nonumber y_1 &=& 12 q_1^{-3/2}  -360 q_1^{-1}  + 39852 q_1^{-1/2}- 829584 + 58417488q_1^{1/2} + \ldots  \,,\\
\nonumber y_2 &=&     12 q_1^{-3/2}  -360 q_1^{-1} - 22356 q_1^{-1/2} + 1617264  -15174576 q_1^{1/2} + \ldots \,,\\
y_3 &=&  36q_1^{-3/2} + 2376 q_1^{-1} -124740 q_1^{-1/2} +3569616  -82184112 q_1^{1/2} + \ldots \,.
\end{eqnarray}
In the large field $\varphi$ approximation, we find that $|y_{1,2,3}|^2 \sim |q_1^{-3/2}|^2 = \exp\left(6\pi e^{\sqrt{\tfrac{1}{3\alpha}}\varphi}\right)$, which implies $\varphi \approx \sqrt{3\alpha}\ln\left(\ln|y_{1,2,3}|^2\right)$. We can construct the modular invariant scalar potential for inflation as follows, taking
$y_1$ as an illustrative example
\begin{equation}\label{eq:vact}
V = V_0\frac{\left[\ln\left( \ln(|y_1(\tau_1)|^2+\beta^2)/\ln(|y_1(\frac{2}{\sqrt{3}}i)|^2+\beta^2)\right)\right]^{2n}}{\left[\ln\left( \ln(|y_1(\tau_1)|^2+\beta^2)/\ln(|y_1(\frac{2}{\sqrt{3}}i)|^2+\beta^2)\right)\right]^{2n} + c}\,,
\end{equation}
where $n$ is a positive integer, $c$ is some positive constant and $\beta$ is a constant to keep logarithm $\ln(|y_1(\tau_1)|^2+\beta^2)$ positive, so that the double logarithm in the potential is well-defined. The extremum of this potential is located at the fixed point $\tau_1 = \frac{2}{\sqrt{3}}i$, a consequence of the Siegel symmetry\footnote{
We now consider the Siegel modular transformation of $\gamma_4$~\cite{Ding:2020zxw}
\begin{equation}
\gamma_4 = \left({\cal S}{\cal T}_3\right)^3\left({\cal S}{\cal T}_1\right)^6{\cal S} =
\begin{pmatrix}
0~&~0~&~0~&~-1\\
0~&~0~&~1~&~0\\
0~&~1~&~0~&~0\\
-1~&~0~&~0~&~0\\
\end{pmatrix}\,.
\end{equation}
Its action on the moduli $\tau$ is
\begin{equation}
\gamma_4\tau = \gamma_4
\begin{pmatrix}
\tau_1 ~&~ \tau_1/2\\
\tau_1/2 ~&~ \tau_1
\end{pmatrix}  =
\begin{pmatrix}
 -\frac{4}{3\tau_1}   ~&~ -\frac{2}{3\tau_1} \\
-\frac{2}{3\tau_1} ~&~-\frac{4}{3\tau_1}
\end{pmatrix} =
\begin{pmatrix}
\tau'_1 ~&~ \tau'_1/2\\
\tau'_1/2 ~&~ \tau'_1
\end{pmatrix} \equiv \tau'\quad \text{with}\quad \tau_1'=-\frac{4}{3\tau_1}\,,
\end{equation}
which implies that $d\tau'_1 = \frac{4}{3}\frac{d\tau_1}{\tau_1^2}$. The potential is modular invariant, thus it fulfills $V(\tau)=V(\tau')$. The derivative of potential with respect to $\tau_1$ at fixed point $\tau_1 = \tau'_1 = \frac{2}{\sqrt{3}}i$ becomes
\begin{equation*}
\frac{d V \left( \tau\right)}{d \tau_1} \bigg|_{\tau_1=\frac{2}{\sqrt{3}}i} = \frac{d V \left( \tau'\right)}{d \tau_1}\bigg|_{\tau_1=\frac{2}{\sqrt{3}}i} =  \frac{4}{3}\frac{1}{\tau_1^2} \frac{d V \left( \tau'\right)}{d \tau'_1} \bigg|_{\tau'_1=\tau_1=\frac{2}{\sqrt{3}}i} = - \frac{d V \left( \tau'\right)}{d \tau'_1} \bigg|_{\tau'_1=\frac{2}{\sqrt{3}}i} = 0\,.
\end{equation*}
}. Obviously, this extremum also is a Minkowski vacuum. In large field approximation, this potential closely resembles the polynomial $\alpha$-attractor model~\cite{Kallosh:2022feu,Bhattacharya:2022akq}, whose inflationary potential and the resulting prediction are given by
\begin{equation}\label{eq:Nens}
V = V_0\frac{\varphi^{2n}}{\varphi^{2n} + c}\,,
\end{equation}
where $c$ is redefined to absorb factor $(3\alpha)^n$ and we see that the potential $V$ approaches $V_0$ in the limit $\varphi \rightarrow \infty$. Similar to the previous scenarios studied in sections~\ref{sec:tau12} and \ref{sec:tau13}, the slope of the potential along axion direction $\theta$ is double-exponentially suppressed, consequently $\theta$ is stabilized during inflation. As an example, we assume it is stabilised to $\theta=0$. Therefore the inflation of the universe is driven by the single inflaton  $\varphi$, in comparison with two field inflation in sections~\ref{sec:tau12} and \ref{sec:tau13}.
\begin{figure}[t!]
\centering
\includegraphics[scale=0.55]{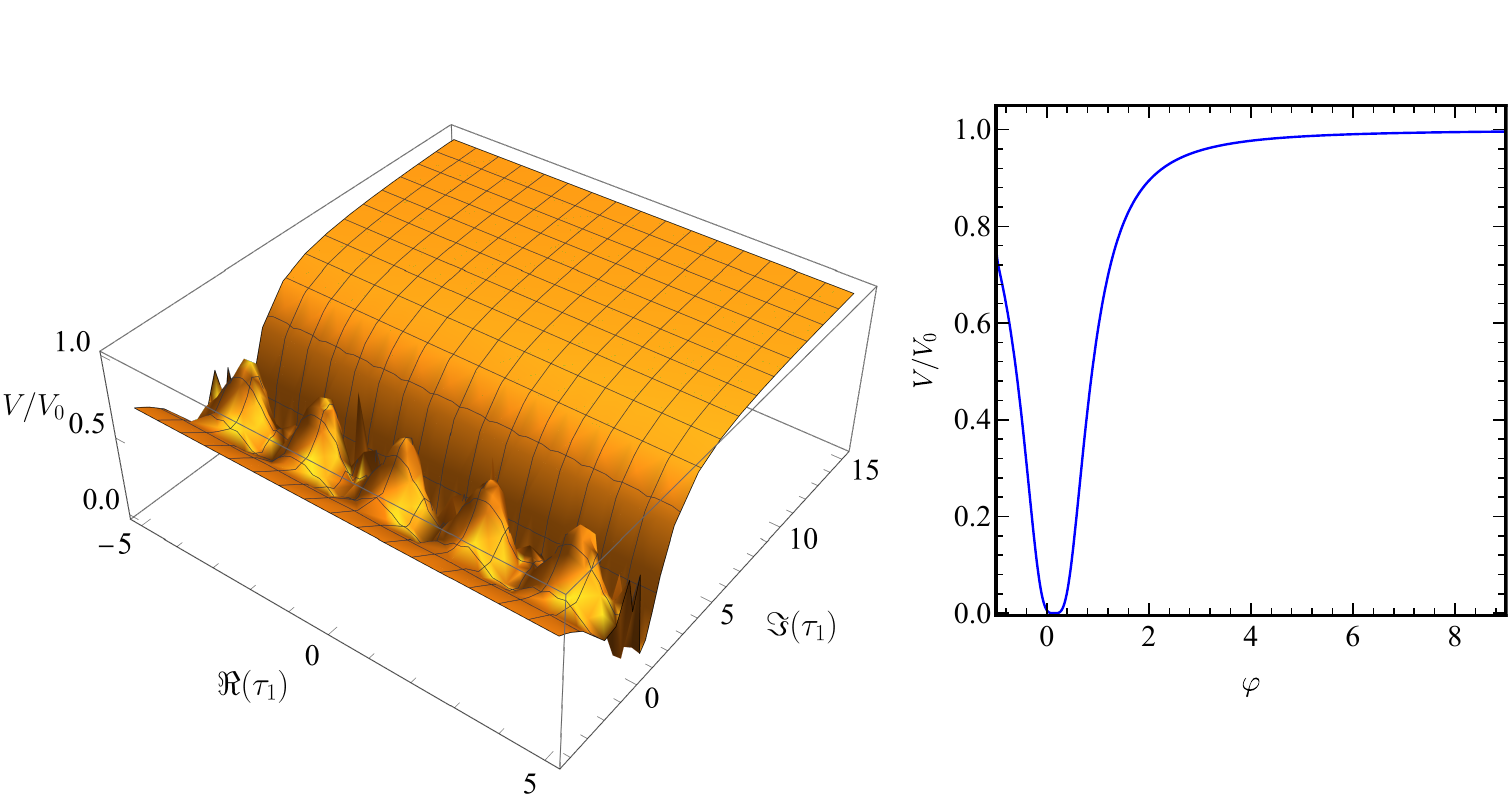}
\caption{The left panel is the variation of the scalar potential $V$ in Eq.~\eqref{eq:vact} with respect to $\Re(\tau_1)$ and $\Im(\tau_1)$, where $\tau_1$ is the complex modulus. The right panel shows this potential as a function of $\varphi$ for $\theta=0$. Here the potential parameters are taken to be $\alpha=1/3$, $\beta=2$ and $c=0.3$.}
\label{fig:plot11dreal}
\end{figure}

In figure~\ref{fig:plot11dreal}, we plot the scalar potential in Eq.~\eqref{eq:vact} with respect to $\Re(\tau_1)$ and $\Im(\tau_1)$ for fixed parameters $n=1$, $\beta=2$ and $c=0.3$. We also present the cross-sectional profile of the potential for $\Re(\tau_1)=\theta=0$ as variation of the single inflaton field $\varphi$ with $\alpha=1/3$. In this scenario, the dotted line in figure~\ref{fig:inflationmodelwhole} shows the $(n_s,r)$ prediction of the single field inflation which lies within the Planck 2018 constrained region~\cite{BICEP:2021xfz}. The predicted tensor-to-scalar ratio $r$ is of order ${\cal O}(10^{-3})$. The model in Eq.~\eqref{eq:vact}, with this choice of parameters, also yields
\begin{equation}
\begin{aligned}
&N = 60\,,\quad n_s = 0.9745\,,\quad r= 0.0017\,,
\end{aligned}
\end{equation}
where the predicted value of $n_s$ remains slightly lower than the analytical estimate $n_s = 0.975$ from Eq.~\eqref{eq:Nens}.  From figure~\ref{fig:inflationmodelwhole}, we can see this point lies slightly outside the $2\sigma$ upper limit of the Planck 2018 observations, yet our prediction in $N=60$ can still fall within the $1\sigma$ region of the P-ACT-LB-BK18\footnote{Here P-ACT-LB-BK18 stands for the combination of the CMB lensing from ACT and Planck and BAO data from DESI DR1 and B-mode measurements from the BICEP and Keck.} dataset~\cite{ACT:2025fju,ACT:2025tim}.

\section{Conclusion and discussions \label{sec:dis}}

The implication of modular symmetry in cosmology has been studied in recent years. Modular symmetry can dictate the interactions between the complex modulus and standard model fields as well as the scalar potential of the modulus. It is notable that a sufficient flat plateau in the potential can be naturally generated to realize the inflation in the Universe. Subsequently the decay of modulus into standard model particles could contribute to the reheating of the Universe~\cite{Ding:2024euc}. Moreover, the cosmological evolution of a CP breaking modulus could generate the baryon asymmetry of the Universe~\cite{Duch:2025abl}. Hence modular symmetry offers a  coherent framework that simultaneously explains crucial open problems in Cosmology.

Modular invariance is a ubiquitous symmetry in string theory, originating from the geometry of compactified extra dimensions. In the simplest torus compactification, the shape of a torus is characterized by the complex structure modulus $\tau$, a dynamical degree of freedom in the effective field theory governed by $SL(2,\mathbb{Z})$ symmetry. Nevertheless, string theory compactification generally yields multiple moduli, requiring a framework that extends beyond the single-field paradigm. The Siegel modular group $Sp(2g,\mathbb{Z})$ serves as a natural generalization of $SL(2,\mathbb{Z})$, capable of accommodating multiple moduli. In the present work, we have considered the Siegel modular group $Sp(4,\mathbb{Z})$ and discussed the inflation driven by the corresponding moduli which is a symmetric complex $2\times 2$ matrix  with positive definite imaginary part.
We focus on specific scenarios involving two-dimensional and one-dimensional complex subspaces. The imaginary components of the moduli are canonically normalized, as derived from the modular invariant K\"ahler potential, and are subsequently identified as the inflaton fields. Guided by these ideas, we use the absolute invariants $y_1$, $y_2$, and $y_3$ of genus $g=2$ which generalize the Klein $j$-invariant of $g=1$ to extend the exponential $\alpha$-attractor to a two-field setup corresponding to the two-moduli subspace. In addition to two-moduli subspaces, we also study the single-modulus case, which allows us to construct modified polynomial $\alpha$-attractor models. We compare the predictions of these models with the Planck 2018 results~\cite{Planck:2018vyg}, finding them all compatible with current observational data.

For the $(\tau_1\,,\tau_2)$ case in section~\ref{sec:tau12}, there exist two independent $SL(2,\mathbb{Z})$ symmetries acting on $\tau_1$ and $\tau_2$, as well as a mirror symmetry under $\tau_1 \leftrightarrow \tau_2$. The absolute invariants $y_1$ and $y_2$ decompose into Klein $j$-invariant $j(\tau_1)$ and $j(\tau_2)$. As an illustrative example, we build an E-model like $\alpha$-attractor potential by using the absolute invariant $y_2$. This potential naturally possesses degenerate Minkowski vacua since $y_2(\tau)=0$ for $\tau_1=i$ or $\tau_2=i$.
For the $(\tau_1\,,\tau_3)$ case in section~\ref{sec:tau13}, the moduli are coupled and have a more intricate structure that requires diagonalization of the K\"ahler metric. We construct a T-model like potential that is sufficiently flat in the large-field limit as well, and realize inflation within the region defined by  $\varphi_2 > \varphi_1 > 0$. Here, since $y_3(\tau)$ vanishes for $\tau_3=0$, this scalar potential has Minkowski vacuum.
In section~\ref{sec:tau11}, we also investigated a single-modulus $Sp(4,\mathbb{Z})$ model where a polynomial $\alpha$-attractor-like potential successfully reproduces the observational predictions of ACT~\cite{ACT:2025fju,ACT:2025tim} and SPT~\cite{SPT-3G:2025bzu} for $N=60$, owing to its ability to accommodate larger values of $n_s$. Meanwhile, the prediction for $N=50$ remains consistent with Planck constraints. Constrained by Siegel modular symmetry, the potential possesses an extremum at the fixed point, which is simultaneously a Minkowski vacuum. The real parts of the moduli in these scenarios are stabilized during inflation, justified by the double exponential suppression of the potential derivative and the validity of the slow-roll approximation. In our model, the inflationary predictions are insensitive to the specific choice of absolute invariants $y_{1,2,3}$ in the large field limit.
It was shown that the $SL(2,\mathbb{Z})$ modular invariant models could be realized in the supergravity~\cite{Kallosh:2024ymt} with the help of a nilpotent superfield~\cite{Komargodski:2009rz,DallAgata:2014qsj,Ferrara:2014kva,Kallosh:2017wnt}. We expect this framework can be straightforwardly extended to the Siegel modular group $Sp(4,\mathbb{Z})$ studied in this work.

Some aspects remain open for future work. We did not address the three moduli case
$(\tau_1\,,\tau_2\,,\tau_3)$
, where canonical normalization and dynamics become considerably more complicated. We also restricted our analysis to the $q$-expansion of absolute invariants without fully exploring their deeper mathematical structure and relationship. This limitation prevented an analytic study of moduli stabilization in our framework. Furthermore, we have not yet considered isocurvature perturbations in these cases, which is an important direction for further study.

\acknowledgments

We acknowledge Dr. Yong Xu for participation in the early stage of this work, reading the paper, providing the data of ACT and professional suggestions.  SYJ and GJD are supported by the National Natural Science Foundation of China under Grant Nos.~12375104, 12547106 and Guizhou Provincial Major Scientific and Technological Program XKBF (2025)010.
WBZ is Supported by the National Natural Science Foundation of China (NSFC) under Grant Nos.~12347103, 12547104 and the Fundamental Research Funds for the Central Universities (grant No.\,E5ER6601A2).


\providecommand{\href}[2]{#2}\begingroup\raggedright\endgroup

\end{document}